\documentclass[journal]{IEEEtran}
\usepackage{makecell}
\usepackage{times}
\usepackage{epsfig}
\usepackage{graphicx}
\usepackage{subfigure}
\usepackage{amsmath}
\usepackage{amssymb}
\usepackage{url}
\usepackage{array}
\usepackage{amsthm}
\usepackage{algorithm}               %format of the algorithm
\usepackage{algorithmic}             %format of the algorithm
\usepackage{multirow}                %multirow for format of table
\usepackage{bm}
\usepackage{footmisc}
\usepackage[pagebackref=true,breaklinks=true,letterpaper=true,colorlinks,bookmarks=false]{hyperref}
\newcommand{\PreserveBackslash}[1]{\let\temp=\\#1\let\\=\temp}
\newcolumntype{C}[1]{>{\PreserveBackslash\centering}p{#1}}
\newcolumntype{R}[1]{>{\PreserveBackslash\raggedleft}p{#1}}
\newcolumntype{L}[1]{>{\PreserveBackslash\raggedright}p{#1}}

\begin{document}

\title{Structural Residual Learning \\for Single Image Rain Removal}

\author{Hong~Wang, Yichen~Wu, Qi~Xie, Qian~Zhao, Yong~Liang,
        and Deyu~Meng,~\IEEEmembership{Member,~IEEE},
       % <-this % stops a space

\thanks{H. Wang, Y. Wu, Q. Xie, Q. Zhao, and D. Meng (corresponding author) are with the Institute for Information and System Sciences and Ministry of Education Key Lab of Intelligent Networks and Network Security, Xi'an Jiaotong University, Shaan'xi, 710049, P.R.China.\,E-mail:\,\{hongwang01,\,wyc620602,\,xq.liwu\}@stu.xjtu.edu.cn;\,\{timmy.zhaoqian,
\,dymeng\}@mail.xjtu.edu.cn.}
\thanks{Y. Liang is with the Faculty of Information Technology, Macau University of Science and Technology, Macau, P.R.China. E-mail:\,yliang@must.edu.mo.}% <-this % stops a space
% <-this % stops a space
}

\markboth{Journal of \LaTeX\ Class Files}%
{Shell \MakeLowercase{\textit{et al.}}: Bare Demo of IEEEtran.cls for IEEE Journals}


\maketitle

% As a general rule, do not put math, special symbols or
%citations in the abstract or keywords.
\begin{abstract}
    To alleviate the adverse effect of rain streaks in image processing tasks, CNN-based single image rain removal methods have been recently proposed. However, the performance of these deep learning methods largely relies on the covering range of rain shapes contained in the pre-collected training rainy-clean image pairs. This makes them easily trapped into the overfitting-to-the-training-samples issue and cannot finely generalize to practical rainy images with complex and diverse rain streaks. Against this generalization issue, this study proposes a new network architecture by enforcing the output residual of the network possess intrinsic rain structures. Such a structural residual setting guarantees the rain layer extracted by the network finely comply with the prior knowledge of general rain streaks, and thus regulates sound rain shapes capable of being well extracted from rainy images in both training and predicting stages. Such a general regularization function naturally leads to both its better training accuracy and testing generalization capability even for those non-seen rain configurations. Such superiority is comprehensively substantiated by experiments implemented on synthetic and real datasets both visually and quantitatively as compared with current state-of-the-art methods.
\end{abstract}

\begin{IEEEkeywords}
    Generalization performance, single image deraining, deep learning, multi-scale learning, convolutional sparse coding, interpretability.
\end{IEEEkeywords}
\section{Introduction}
\IEEEPARstart{I}{mages} captured in rainy days often suffer from noticeable degradation of scene visibility and adversely affect the performance of subsequent image processing tasks, such as video surveillance~\cite{Shehata2008Video,bahnsen2018rain} and intelligent vehicles~\cite{SmithVisual}. Removing rain streaks from a rainy image thus has become a necessary step for a wide range of practical applications, and has attracted much research attention recently~\cite{benchmark,yasarla2020confidence}.

\begin{figure}[htb]
    \centering
        \includegraphics[width=1\linewidth]{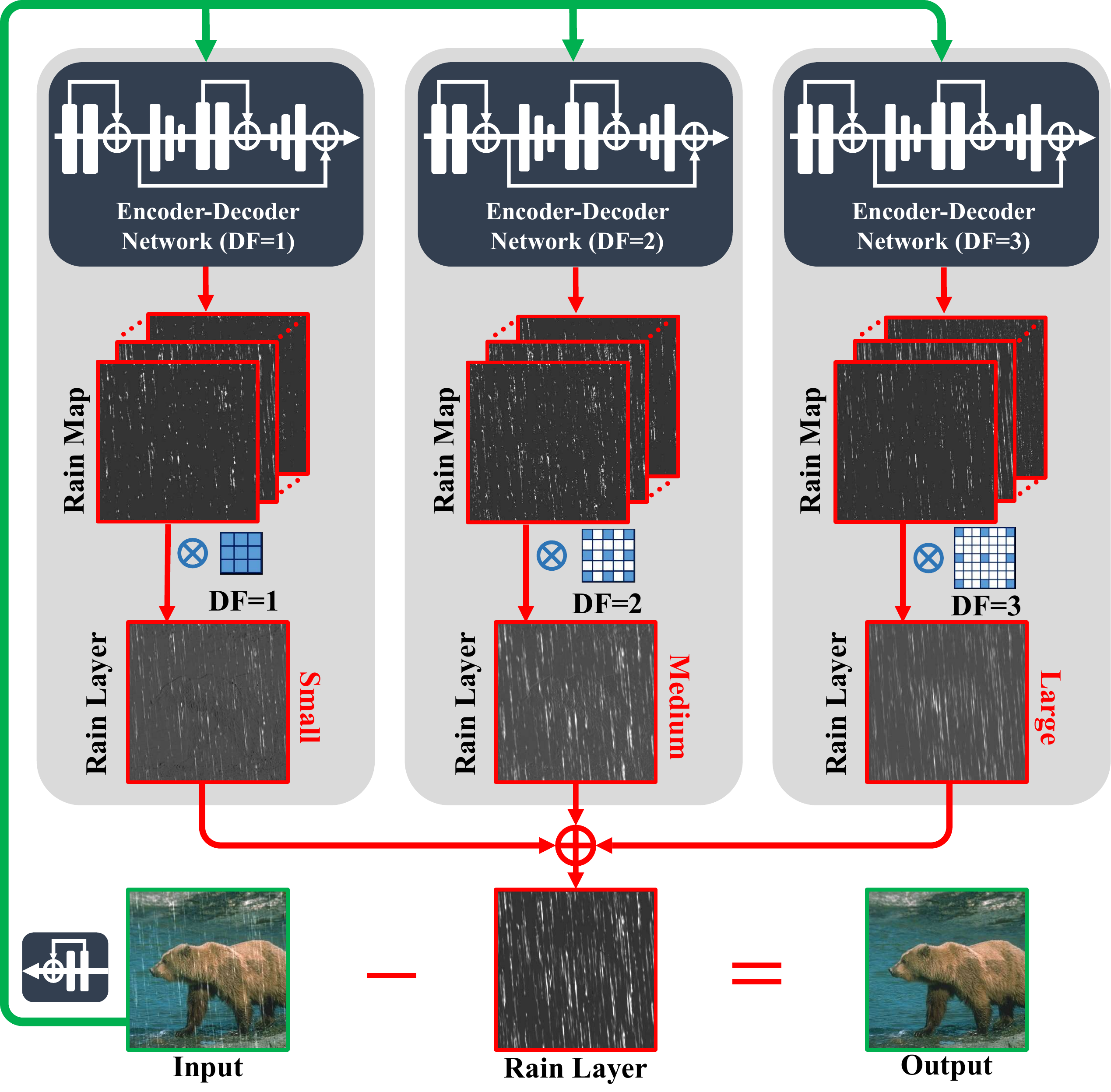}%\vspace{-1mm}
    \caption{Illustration of the proposed structural residual network (SRNet), which consists of three parallel subnetworks with similar structures but different dilated factors (DF) to capture rational structures expressing rain streaks.}
    \label{intro_net}
      \vspace{-1.2mm}
\end{figure}

Many single image deraining methods have been raised in the recent decades. Some early attempts exploited different filters to decompose a rainy image into low frequency part (LFP) and high frequency part (HFP), such as guided filter~\cite{Jing2012Removing,He2010Guided} and $L_0$ smoothing filter~\cite{Ding2016Single}, and then restored the rain-free image by combining the LFP and texture details.
Recently, many researchers further explored the physical properties of rain and background layers and formulated them into different prior terms for single image rain removal. For example, the techniques of Gaussian mixture model (GMM)~\cite{Li2016Rain} and discriminative sparse coding (DSC)~\cite{Yu2015Removing} have been used to model rain layers. Besides, joint convolutional analysis and synthesis sparse representation (JCAS)~\cite{Gu2017Joint} is a typical instance that describes rain and background layers with concise mathematical models. These methods have been shown to be effective in certain scenarios. Such prior-model-based techniques, however, are still lack of the ability to flexibly adapt to rainy images with complicated rain shapes and background scenes. Besides, these methods are generally time-consuming due to their inevitable iterative optimization computations, which is always unfriendly to real users.

Very recently, driven by the success of deep learning (DL) techniques in low level vision tasks, the related techniques have also been employed in the deraining task. They are routinely required to pre-collect abundant training samples with rainy-clean image pairs to learn non-linear mapping from rainy image to its rain-free one in an end-to-end manner. The most representative methods along this research line include DerainNet~\cite{Fu2017Clearing}, deep detail network (DDN)~\cite{Fu2017Removing}, deep joint rain detection and removal network (JORDER\_E)~\cite{yang2017deep,Yang2019Joint}, recurrent squeeze-and-excitation context aggregation network (RESCAN)~\cite{li2018recurrent}, progressive image deraining network (PReNet)~\cite{ren2019progressive}, and spatial attentive network (SPANet)~\cite{wang2019spatial}.

Albeit achieving success in certain contexts, the current DL methods for single image deraining still exist evident limitations. Specifically, the effectiveness of these methods largely relies on the quality and quantity of pre-collected training samples which are composed of large amount of rainy-clean images simulating the network inputs and outputs. The training images should include possibly wide range of rain shapes so as to cover those potentially occurring in testing stages. On the one hand, the non-linear mapping represented by the deep network architecture brings strong fitting capability on approximating abundant rain structures contained in training images beyond conventional model-based approaches. On the other hand, however, their complex network expression also tends to conduct redundancy in expressing rain streaks in training samples and excessive flexibility on adapting to newly input rainy images. A DL model achieving good training accuracy might not finely extract rains (like those containing evident background scenes, as shown in our experiments) from some practical rainy images with complex rain streaks non-seen in training stages. How to alleviate such generalization issue has become the core issue nowadays in current single image deraining research~\cite{wang2019spatial}.

%\begin{figure}
%    \centering
%        \includegraphics[width=1\linewidth]{intro.pdf}%\vspace{-1mm}
%    \caption{Generalization performance comparison of different single image derainers. (a) Real rainy image. (b)-(d) The derained results of SPANet, JORDER\_E, and the proposed SRNet, respectively.}
%    \label{intro_fan}
%%\vspace{-2mm}
%\end{figure}

Against this issue, we build a specific network architecture, called structural residual network (SRNet), by enforcing its output residual finely accordant with the prior expressions for general rains. Such a structural residual setting forms a strong regularizer for rain layer extraction, guaranteeing the rationality of output rain streaks even for non-seen rain shapes differentiated from those contained in training samples, and is thus expected to achieve generalization ability in testing stages.

Specifically, inspired by the previous research on rain structure modeling, we summarize the following prior knowledge for representing general rain streaks, and employ them for building our network architecture. Firstly, as explored in~\cite{Gu2017Joint} and ~\cite{He2017Convolutional}, rain streaks always repeatedly appear at different locations over a rainy image with similar local patterns like shape, thickness, and direction. Such knowledge can be delivered by convolutional representation under a set of local filters across the entire images. The structures of these filters are expected to capture local repetitive patterns inside rains. Secondly, an evident prior of rains is that they are always sparsely scattered over an image~\cite{Yu2015Removing,Zhu2017Joint}. This knowledge can be expressed by the sparsity of feature maps (can be seen as coefficients) convoluted with those local filters. In our network, each feature map of the last layer is yielded by an embedded encoder-decoder subnetwork, in which the joint MaxPooling and MaxUnpooling operation naturally leads to the sparsity of all last-layer feature maps. Thirdly, another common prior used for describing rain streaks are their multi-scale characteristics~\cite{li2018video}, referring to the fact that rain streaks in different distances appear with different sizes in a rainy image. We thus construct the output of our network as an integration of multiple  encoder-decoder subnetworks through utilizing dilated convolutions~\cite{yu2015multi} with different dilated factors. This facilitates a rational expression for extracted rain layers with different scales. The architecture is demonstrated in Fig. \ref{intro_net} for easy observation.

Note that such a structural residual setting keeps the rain layer extracted by the network comply with the prior knowledge underlying general rains, and thus is functioned like a conventional regularizer to rectify output rain shapes being well extracted from rainy images by the network in both training and predicting stages. This naturally leads to both its potential good training accuracy and testing generalization capability even for those non-seen rain configurations. Such superiority of the proposed network is substantiated by comprehensive experiments implemented on synthetic and real datasets, especially on its fine generalization capability. Furthermore, ablation studies have been provided to verify the necessity of all modules involved in our network.

The paper is organized as follows. Section \uppercase\expandafter{\romannumeral2} briefly reviews related works. Section \uppercase\expandafter{\romannumeral3} presents the proposed structural residual single image deraining network as well as the network training details. Comprehensive experiments are shown in Section IV and the paper is finally concluded.

\section{Related Work}
In this section, we briefly review current development along video deraining methods and single image deraining methods.

\subsection{Video Deraining Methods}
Garg and Nayar~\cite{Garg2004Detection,gk_vision} initially discussed the visual effects of raindrops on imaging systems and proposed a video deraining method, by capturing dynamics of raindrops through a space-time correlation model and describing the photometry of rain via a motion blur model. Later, researchers made more investigations on exploring physical properties of rain streaks for video deraining, like temporal-chromatic~\cite{zhang2006rain, park2008rain} and spatio-temporal frequency characteristics~\cite{barnum2010analysis,tripathi2012video}.

In the past years, more intrinsic prior structures of rain streaks and background scenes of rainy videos have been analyzed and formulated for deraining algorithm design. For example, Chen~\emph{et al.}~\cite{Chen2013A} investigated the non-local similarity and repeatability of rain streaks and provided a low rank model. To remove rain and snow from a video, Kim~\emph{et al.}~\cite{Jin2015Video} adopted a low rank matrix completion strategy. To handle heavy rain streaks and dynamic scenes, Ren~\emph{et al.}~\cite{Ren2017Video} decomposed rain streaks into two classes: sparse ones and dense ones, and detected them with different models. Recently, Wei~\emph{et al.}~\cite{wei2017should} used stochastic manner to encode rain streaks as a patch based mixture of Gaussian. Li~\emph{et al.}\cite{li2018video} further explored the prior structure of rain streaks and described that they have the repetitive local patterns and multi-scale characteristics. Based on the observation, the authors proposed a multi-scale convolutional sparse coding model that achieves the state-of-the-art deraining performance when the background layer in a video is also finely extracted.

Very recently, deep learning techniques have achieved great success in various low-level vision tasks, such as image denoising~\cite{zhang2017beyond,zhang2018ffdnet}, image super-resolution~\cite{dong2015image,zhang2018residual,tai2017image}, and image deblurring~\cite{tao2018scale,kupyn2018deblurgan}. Recent years have also witnessed the development of deep learning in the deraining task~\cite{Jie2018Robust}. To remove rains from videos, Liu~\emph{et al.}~\cite{liu2018erase} provided a hybrid rain model and constructed a multi-task learning network architecture that successively accomplished rain degradation classification, rain removal, and background restoration. Further, the authors developed a dynamic routing residue recurrent network~\cite{Liu2018D3R} for handling dynamically detected rainy videos. Compared with these video deraining methods, the single image deraining task is much more challenging in practice due to the lack of temporal information.

\subsection{Single Image Deraining Methods}
Against this single image deraining task, Xu~\emph{et al.}~\cite{Jing2012Removing} considered the chromatic property of rain streaks and achieved a coarse rain-free image via a guided filter. Kim~\emph{et al.}~\cite{Kim2014Single} analyzed the geometric property of rains to detect rain streak regions, and then reconstructed the derained result by executing nonlocal means filtering on the detected region.

Later on, researchers resorted to utilizing domain knowledge to encode rains for helping the deraining task~\cite{Kang2012Automatic,Wang2017A}. For example, based on morphological component analysis, Fu~\emph{et al.}~\cite{fu2011single} considered single image deraining as a signal decomposition problem. Then the authors executed a bilateral filter, dictionary learning, and sparse coding to acquire LFP and HFP for obtaining the derained result. Afterwards, Luo~\emph{et al.}~\cite{Yu2015Removing} adopted a screen blend model and proposed to utilize high discriminative codes  over a learned dictionary to sparsely approximate rain and background layers. To represent multiple orientations and scales of rain streaks, Li~\emph{et al.}~\cite{Li2016Rain} designed GMM based patch prior. By employing the sparsity and gradient statistics of rain and background layers, Zhu~\emph{et al.}~\cite{Zhu2017Joint} formulated three regularization terms to progressively extract rain streaks. Afterwards, Gu~\emph{et al.}~\cite{Gu2017Joint} utilized analysis sparse representation to represent image large-scale structures and synthesis sparse representation to describe image fine-scale textures. Meanwhile, Zhang~\emph{et al.}~\cite{He2017Convolutional} proposed to learn a set of sparsity based and low rank based convolutional filters for illustrating background and rain layers, respectively. Albeit achieving good performance on certain scenarios, these prior-model-based methods need inevitable time-consuming iterative inferences and always could not finely fit practical diverse rain shapes in real rainy images due to relatively simple while subjective prior assumptions.

More recently, some CNN based DL methods have been raised for the task~\cite{li2018recurrent,zhang2019image,wang2019spatial}. Fu \emph{et al.}~\cite{Fu2017Clearing} first designed the DerainNet to predict clean image. To make training process easier, the authors~\cite{Fu2017Removing} further proposed DDN to remove rain content in HFP. Later, Zhang~\emph{et al.}~\cite{zhang2019image} proposed a conditional generative adversarial deraining network. Considering the complicatedness of rain streaks, the authors~\cite{zhang2018density} further incorporated a residual-aware classifier process and designed a rain density-aware multistream dense network. Furthermore, Yang~\emph{et al.}~\cite{yang2017deep} reformulated the commonly-used rain model and developed a multi-task architecture to jointly learn binary rain streak map, appearance of rain streaks, and clean background. To achieve better visual quality, the authors further introduced a detail preservation step~\cite{Yang2019Joint}. Very recently, by repeatedly unfolding a shallow ResNet with a recurrent layer, Ren~\emph{et al.}~\cite{ren2019progressive} presented a simple baseline network, PReNet. To make the network better reflect the prior knowledge of rain structures, some beneficial attempts have also been made. For example, Mu~\emph{et al.}~\cite{Mu2019Learning} proposed to implicitly describe prior structures by data-dependent networks and then formulate them into bi-layer optimization iterations. To alleviate the hard-to-collect-training-sample and overfitting-to-training-sample issues, Wei \emph{et al.}~\cite{wei2019semi} formulate rain layer prior as GMM and train the backbone, DDN, in a semi-supervised manner. Albeit good performance, these DL methods embed these beneficial prior rain knowledge inside the network architecture~\cite{Mu2019Learning} or on the loss formulation for training the network~\cite{wei2019semi}. The high flexibility of the nonlinear network mapping might output unexpected mappings deviated from practical rain shapes especially in the testing stages with non-seen rain configurations in training samples. Their generalization performance could thus possibly be negatively influenced. This is the main motivation of this study, to make network extract rain layer, possibly complying with the intrinsic rain structures the previous researches have been explored, from a rainy image.

\begin{figure*}
    \centering
    \includegraphics[width=1\linewidth]
       {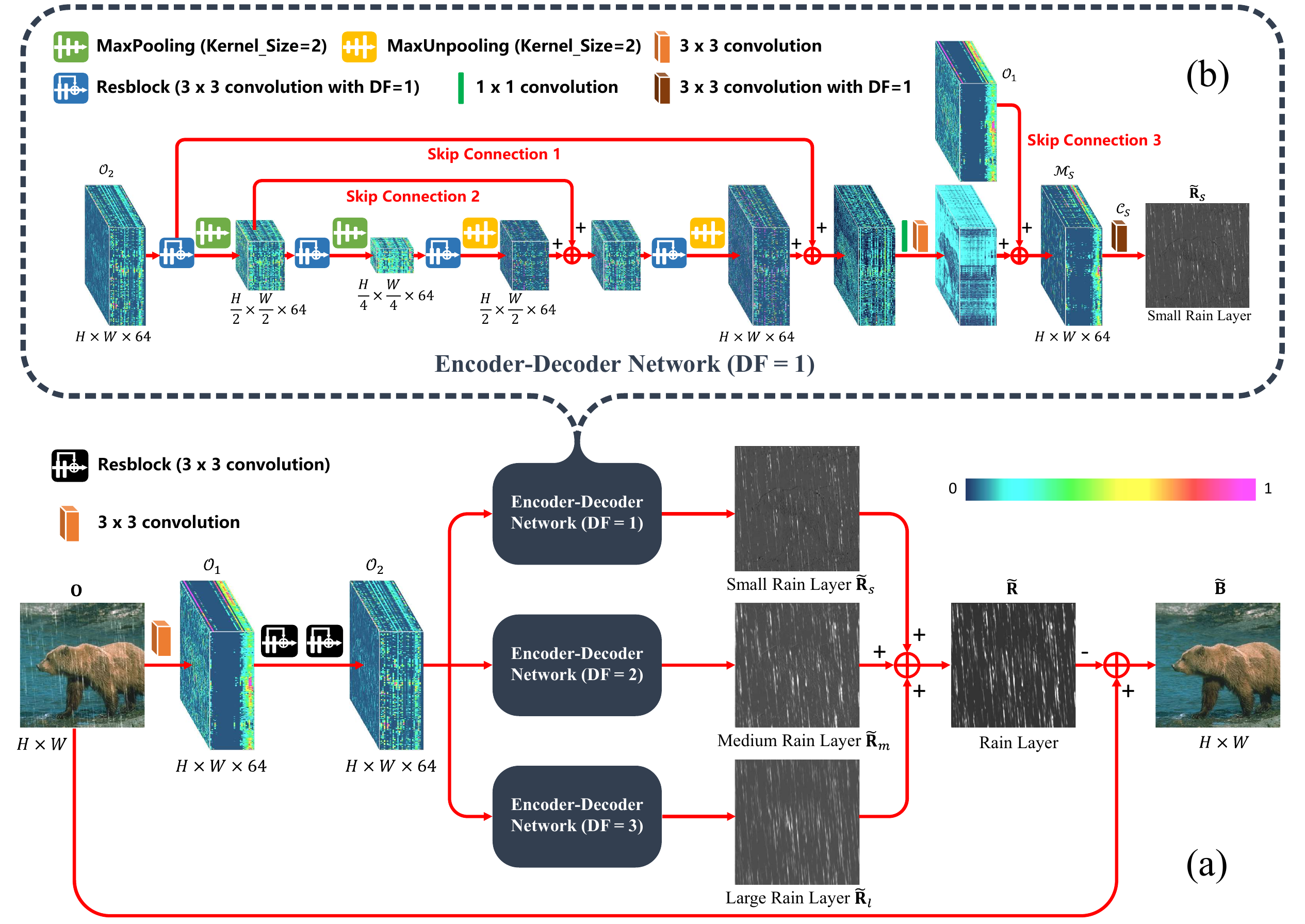}%\vspace{-1mm}
    \caption{Illustration of the proposed SRNet. (a) The whole deraining network architecture. (b) The Encoder-Decoder Network for extracting rain layer $\widetilde{\mathbf{R}}_{s}$ at relatively small scale. The skip connections 1 and 2 achieve local residual learning and the skip connection 3 denotes global residual learning~\cite{tai2017image}. Besides, local residual learning also includes the skip connection in each Resblock~\cite{he2016deep}. For obtaining relatively medium rain layer $\widetilde{\mathbf{R}}_{m}$ and large rain layer $\widetilde{\mathbf{R}}_{l}$, the dilated factor $DF=1$ in (b) is replaced with 2 and 3, respectively.
    %For better understanding, we have utilized the pseudo-color way to visualize the intermediate outputs.
    }
    \label{mseda}
%\vspace{-2mm}
\end{figure*}

\section{Structural Residual Network for Single Image Rain Removal}

\subsection{Basic Rainy Image Model}
An input rainy image is denoted as $\mathbf{O}\in\mathbb{R}^{H\times W}$, where $H$ and $W$ denote its height and width, respectively. The commonly-used rainy image model is:
\begin{equation}\label{e1}
\mathbf{O}=\mathbf{B}+\mathbf{R},
\end{equation}
where $\mathbf{B}$ and $\mathbf{R}$ represent the background layer and the rain layer of the image, respectively. The goal of DL based single image derainers is to design rational network architectures to learn non-linear mapping functions from an input rainy image $\mathbf{O}$ to its background layer $\mathbf{B}$ or the residual rain layer $\mathbf{R}$.

Our aim is to construct a network architecture with its output residual capable of sufficiently representing prior rain structures explored by previous investigations, so as to rectify a rational and interpretable network prediction even for those non-seen and complicated rainy image inputs. Specifically, three known priors have been considered in this network design task. The first is the local-repetitive-pattern prior as proposed in~\cite{Gu2017Joint} and ~\cite{He2017Convolutional}. That is, rain streaks always repeatedly appear at different locations over a rainy image with similar local patterns like shape, thickness, and direction. The second is the coefficient-sparsity prior, which describes the fact that the rain streaks are always sparsely scattered over an image~\cite{Yu2015Removing,Zhu2017Joint}. The third is the multi-scale prior~\cite{li2018video}, which refers to the phenomenon that rain streaks in different distances appear with different sizes in a rainy image since they are pictured from different distances by cameras. Based on this, the single image rainy model (\ref{e1}) can be more elaborately formulated as:
\begin{equation}\label{e2}
\mathbf{O}= \mathbf{B}+  \mathbf{R}_{s} + \mathbf{R}_{m} +  \mathbf{R}_{l},
\end{equation}
%\end{small}
where $\mathbf{R}_{s}$, $\mathbf{R}_{m}$, and $\mathbf{R}_{l}$ denote three different scales of rain layer separation, respectively.

We then introduce how to construct a deep network with carefully designed structural residual output finely conveying all the aforementioned prior knowledge of rain streaks.

\subsection{Structural Residual Network Architecture}
In this section, we construct a single image deraining network to learn structural residual output. As shown in Fig.~\ref{mseda}, the proposed network, called structural residual network (SRNet), mainly consists of three parallel sub-networks with similar structures but different dilated factors (DF), to extract rain layer at different scales. The rain-removed result $\widetilde{\mathbf{B}}$ is estimated by subtracting the entire residual rain layer $\widetilde{\mathbf{R}}$ from the input $\mathbf{O}$. The design details are described as follows.

\begin{figure}[t]
    \centering
        \includegraphics[width=1\linewidth]{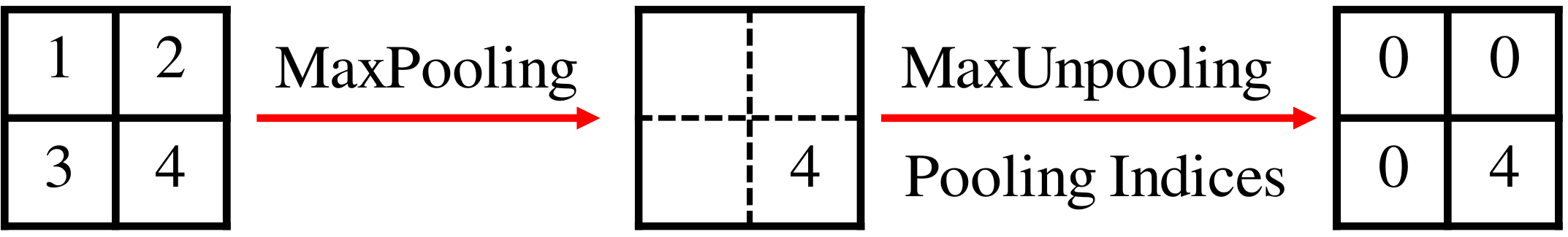}%\vspace{-1mm}
    \caption{Illustration of the pairwise MaxPooling and MaxUpooling operation with kernel\_size as 2\cite{paszke2017automatic}.}
    \label{pool}
   % \vspace{-3mm}
\end{figure}
\begin{figure}[t]
    \centering
        \includegraphics[width=1\linewidth]{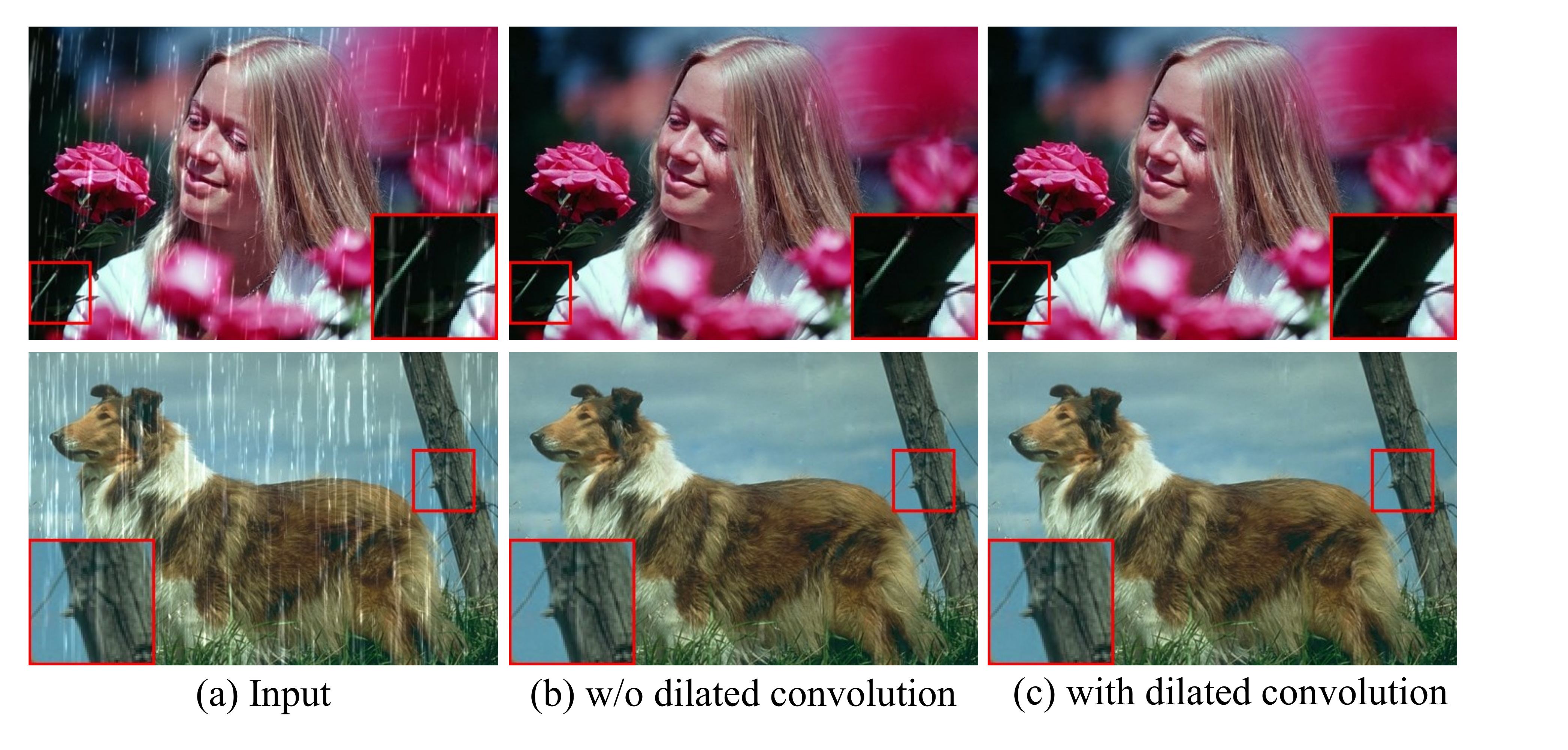}%\vspace{-2mm}
    \caption{Rain removal performance comparison on two typical rainy images. (a) Original rainy images. (b) The derained results in the case of ``without dilated convolution'' (only with small branch). (c) The derained results in the case of ``with dilated convolution''.}
    \label{dilat}
      \vspace{-2mm}
\end{figure}
\begin{figure*}[t]
    \centering
        \includegraphics[width=1.01\linewidth]{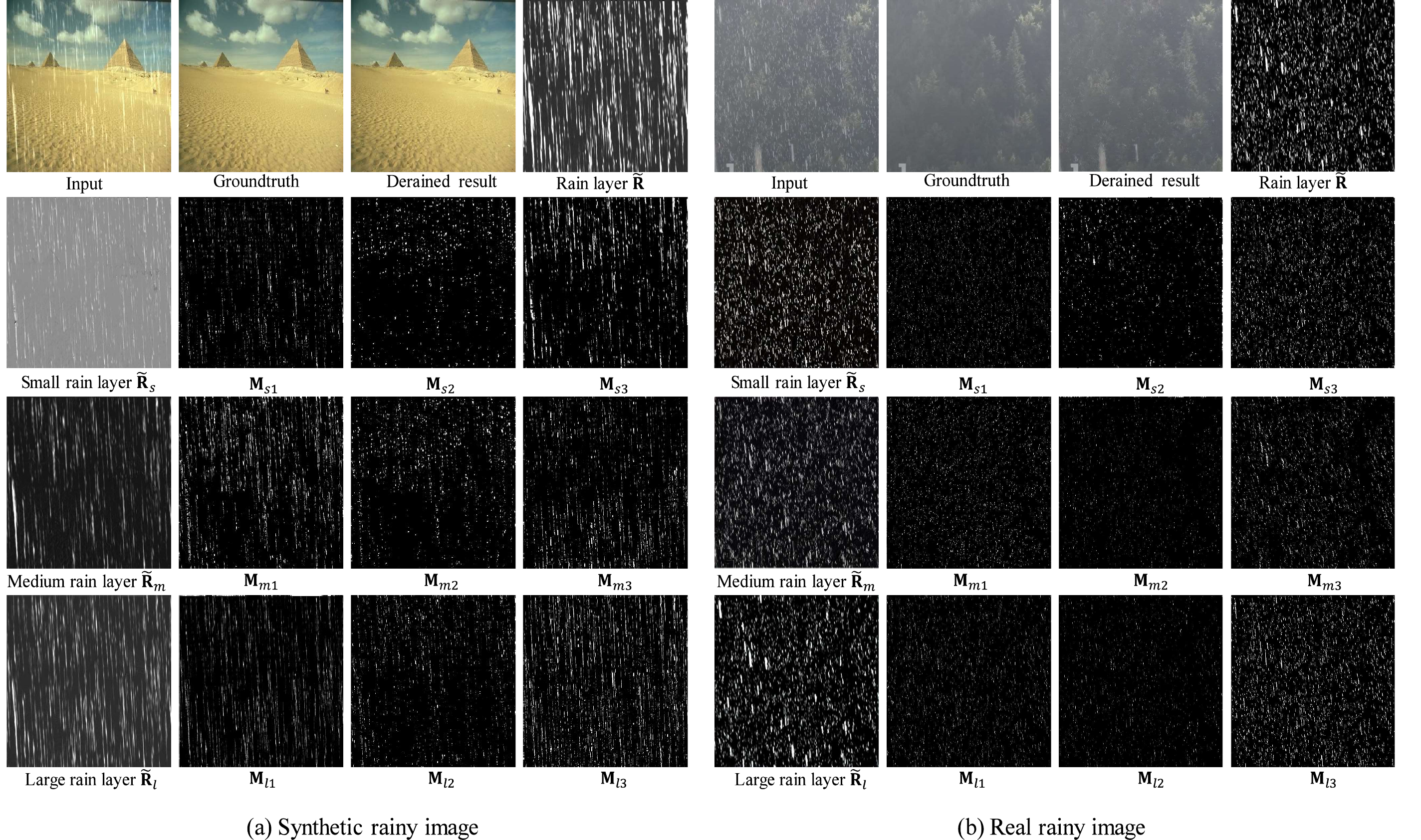}%\vspace{-2mm}
%    \caption{Structured rain layer visualization for (a) a synthetic rainy image from Rain100L and (b) a real rainy image from SPA-Data. Taking (a) as an exmaple, the first row from left to right: rainy image, groundtruth, derained result, and rain layer $\widetilde{\mathbf{R}}$ extracted by MSEDNet. The second row: rain layer at small scale $\widetilde{\mathbf{R}}_{s}$ and the corresponding three typical sparse rain feature maps, i.e., $\mathbf{M}_{s1}$, $\mathbf{M}_{s2}$, and $\mathbf{M}_{s3}$. The third row: rain layer at medium scale $\widetilde{\mathbf{R}}_{m}$ and the corresponding three typical sparse feature maps, i.e., $\mathbf{M}_{m1}$, $\mathbf{M}_{m2}$, and $\mathbf{M}_{m3}$. The forth row: rain layer at large scale $\widetilde{\mathbf{R}}_{l}$ and the corresponding three typical sparse feature maps, i.e., $\mathbf{M}_{l1}$, $\mathbf{M}_{l2}$, and $\mathbf{M}_{l3}$.}
    \caption{Structural rain layers extracted by the SRNet for (a) a synthetic rainy image from Rain100L~\cite{Yang2019Joint} and (b) a real rainy image from SPA-Data~\cite{wang2019spatial}. In both panel, the first row denote (from left to right): rainy image, groundtruth, derained result, and extracted rain layer $\widetilde{\mathbf{R}}$. The second to forth rows denote: the first column is the output rain layers $\widetilde{\mathbf{R}}_{s}$, $\widetilde{\mathbf{R}}_{m}$, and $\widetilde{\mathbf{R}}_{l}$ obtained by our method; The last three columns are the corresponding three representative rain feature maps, i.e., $\mathbf{M}_{*1}$, $\mathbf{M}_{*2}$, and $\mathbf{M}_{*3}$ at the scale $*$ (*$\in \{s, m, l\}$). The figure is better observed by zooming in on a computer screen.}
    \label{visnet}
  %  \vspace{-2mm}
\end{figure*}
\subsubsection{Convolutional Sparse Coding Structure for Rain Streaks}
As displayed in Fig.~\ref{mseda}(a), the SRNet first adopts one convolution layer and two Resblocks to extract shallow features $\mathcal{O}_{2}\in \mathcal{R}^{H \times W \times N}$. Consistent to  the  model (\ref{e2}), three encoder-decoder subnetworks are constructed by exploiting  the dilated convolutions~\cite{yu2015multi} with different DFs to extract $\widetilde{\mathbf{R}}_{s}$, $\widetilde{\mathbf{R}}_{m}$, and $\widetilde{\mathbf{R}}_{l}$ at relatively small, medium, and large scales, respectively. Here we take the extraction of small rain layer $\widetilde{\mathbf{R}}_{s}$ as an example and illustrate the encoder-decoder network structure.

As shown in Fig.~\ref{mseda}(b), the proposed encoder-decoder subnetwork is symmetrically composed of encoder and decoder parts. In the encoder part, two consecutive Resblocks are to obtain deeper features of the shallow feature input $\mathcal{O}_{2}$. Each Resblock is followed by one MaxPooling layer with downsampling ratio set as $2$. Symmetrically, two extra Resblocks are stacked in the decoder part and each one is followed by one MaxUnpooling layer for upsampling the feature activations ahead. The MaxUnpooling layer exploits the pooling indices computed on the corresponding MaxPooling layer to execute non-linear unsampling, as shown in Fig.~\ref{pool}. With the symmetrical combination of MaxPooling/MaxUnpooling layers, and the feature fusion operation with successive $1\times1$ convolution and $3\times 3$ convolution, we get the sparse rain feature map $\mathcal{M}_{s}\in \mathcal{R}^{H \times W \times N }$, as displayed in Fig.~\ref{mseda}(b). Such sparsity finely represents the aforementioned coefficient-sparsity prior, i.e., the sparse locations of rain streaks across the image.
Subsequently, by imposing the convolution layer $\mathcal{C}_{s}$ on $\mathcal{M}_{s}$, small rain layer $\widetilde{\mathbf{R}}_{s}$ can then be extracted. Such convolutional coding manner naturally delivers the local-repetitive-pattern prior of rain shapes over the entire image.

From \cite{zhang2018residual} and \cite{huang2017densely}, incorporating skip or dense connection strategy into a deeper model can improve the understanding and representation of image contents. Hence, to fully utilize the shallow feature $\mathcal{O}_{1}$ and propagate the long-distance spatial context information, we introduce the global residual learning~\cite{tai2017image} into the encoder-decoder network shown as skip connection 3 in Fig.~\ref{mseda}(b). Besides, we incorporate diverse local residual learning, including the skip connection inside each Resblock and the skip connections between the encoder part and the decoder part shown as 1 and 2 in Fig.~\ref{mseda}(b). In this way, we aim to deepen into stronger feature expressions for better extracting rain structure. Meanwhile, the training process would be made easier~\cite{he2016deep}. The functions of such MaxUnpooling layer and skip connection settings to the final performance of our network will be validated in our following ablation experiments.
\begin{figure*}[t]
    \centering
        \includegraphics[width=1\linewidth]{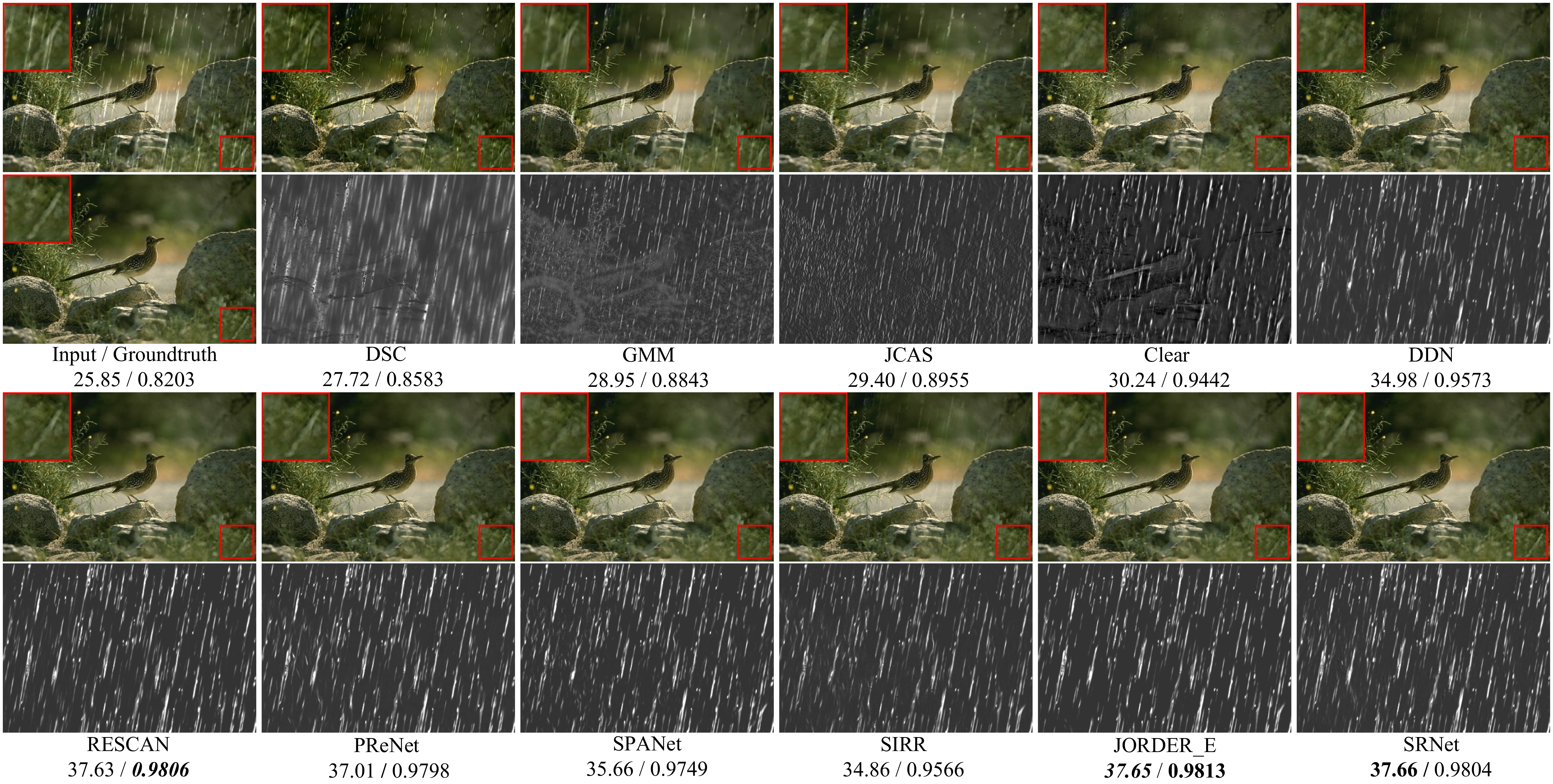}%\vspace{-2mm}
    \caption{Performance comparison of all competing methods on a typical test image in Rain100L. The first row: rainy image and derained results obtained by different methods. The second row: groundtruth and rain layers extracted by different methods. PSNR/SSIM results are given below each figure for easy reference. Bold and bold italic indicate top $1^{\text{st}}$ and $2^{\text{nd}}$ best performance, respectively.}
    \label{100l}
    %    \vspace{-3mm}
\end{figure*}
\subsubsection{Multi-Scale Structure for Rain Streaks}
To represent the multi-scale configuration of rain streaks, we simply resort to dilated convolution, which weights pixels with a step size of DF, and enlarges its receptive field without losing resolution~\cite{yu2015multi}. Specifically, we repeatedly construct the proposed encoder-decoder subnetwork with the same kernel size $3 \times 3$ but with three different DFs (1, 2, and 3), as displayed in Fig.~\ref{mseda}(a). The three branches thus have their own receptive fields to capture information at different scales. Fig.~\ref{dilat} demonstrates the effect of this multi-scale specification by comparing the rain-removed images restored by our network with and w/o dilated convolutions. One can easily observe that with dilated convolution, the SRNet can acquire larger receptive field that inclines to help remove long rain streaks as well as preserve large-scale image details.

Here, we present a visualization for the structural residual learned by the poposed SRNet to express the rain layer as shown in Fig.~\ref{visnet}. The rain layer and rain feature maps ($\mathbf{M}_{*1}$, $\mathbf{M}_{*2}$, and $\mathbf{M}_{*3}$) on two testing images by our network are demonstrated. It is easy to see obvious differences among $\widetilde{\mathbf{R}}_{s}$, $\widetilde{\mathbf{R}}_{m}$, and $\widetilde{\mathbf{R}}_{l}$ with multi-scale characteristics. $\mathbf{M}_{*1}$, $\mathbf{M}_{*2}$, and $\mathbf{M}_{*3}$ approximately illustrate the directions, positions, and shapes of rain streaks, implying the convolutional coding mechanism of the proposed method. In such manner, the proposed SRNet is expected to capture the rational rain streaks complying with these pre-known prior structures.
\subsection{Objective Function}
%According to Fig.~\ref{msed} and Fig.~\ref{pool}, it is easy to derive our network structure mathematically. For brevity, here we only provide some key formulas as:
%\begin{equation}\label{net}
%\begin{split}
%&f^{d}(\mathbf{O}) = \text{max}(0,\mathbf{W}^{d-1}\otimes f^{d-1}(\mathbf{O})+\mathbf{b}^{d-1}), d = 1,2\\
%&\tilde{\mathbf{R}}_{s}= \mathbf{C}_{s}\otimes  \mathbf{M}_{s}, ~~~\\&\widetilde{\mathbf{R}}_{m}= \mathbf{C}_{m}\otimes  \mathbf{M}_{m},\\
%&\tilde{\mathbf{R}}_{l}= \mathbf{C}_{l}\otimes  \mathbf{M}_{l},~~~\\&\widetilde{\mathbf{B}}=\mathbf{O}-(\widetilde{\mathbf{R}}_{s}+\tilde{\mathbf{R}}_{m}+
%\tilde{\mathbf{R}_{l}}),
%\end{split}
%\end{equation}
%where $\tilde{\mathbf{B}}$ is the derained result, and $\tilde{\mathbf{R}}_{s}$, $\tilde{\mathbf{R}}_{m}$, and $\tilde{\mathbf{R}}_{l}$ are extracted rain layers at three scales. $\mathbf{W}^{d-1}$ and $\mathbf{b}^{d-1}$ are learnable network parameters. We define $f^{0}(\mathbf{O})=\mathbf{O}$.  The first recursive formula in (\ref{net}) is a general expression representing convolution layer and $d$ is the recursive times. In specific, in the case of $d=1$, it only means a convolution layer, for example, $\mathbf{O}_{1}=\text{max}(0,\mathbf{W}^{0}\otimes  \mathbf{O}+\mathbf{b}^{0})$. With $d=2$, it denotes a Resblock consisting of two consecutive convolution layers.
As the sensitivity of the human visual system depends on local luminance, contrast, and structure, this can be quantitatively measured by the structure similarity (SSIM) index~\cite{Zhou2004Image}. The larger the SSIM value is, the better the quality of the image is. To train the SRNet, similar to~\cite{ren2019progressive}, we simply adopt the negative SSIM as the objective function, written as:
\begin{equation}
\mathcal{L}=-\text{SSIM}\left({\widetilde{\mathbf{B}}}, \mathbf{B}\right).
\end{equation}
%where $\mathbf{B}$ is the groundtruth.
\subsection{Training Details}
We use PyTorch\cite{paszke2017automatic} to implement the proposed SRNet, based on a PC equipped with an Intel (R) Core(TM) i7-8700K at 3.70GHZ and one Nvidia GeForce GTX 1080Ti GPU. Since we mainly care about the effectiveness of the network architecture on rain removal, instead of relying on some extra tricks such as designing complicated loss functions and tuning hyper-parameters, we directly adopt the parameter settings of the latest baseline network--PReNet~\cite{ren2019progressive}. Specifically, the patch size is 100$\times$100 and the Adam optimizer\cite{Kingma2014Adam} with the batch size of 18 is used. The initial learning rate is $1\times$$10^{-3}$ and divided by 5 after reaching 30, 50, and 80 epochs. The total epoch is 100. It is worth mentioning that for all datasets in subsequent experiments, these parameter settings are the same. This would show the favorable robustness and generality of our method. Besides, to balance the trade-off between running time and deraining performance on synthetic and real datasets, we simply select depth $T=2$ and width $N=64$ as the default setting in our experiments, where the depth $T$ denotes the times of symmetrical downsampling/upsampling operations, namely, the number of the module Resblock+MaxPooing (Resblock+MaxUnpooing) in Fig.~\ref{mseda}(b).
%All the related source codes will be released in http://dymeng.gr.xjtu.edu.cn.
\begin{figure*}[t]
    \centering
   % \vspace{-4mm}
        \includegraphics[width=1\linewidth]{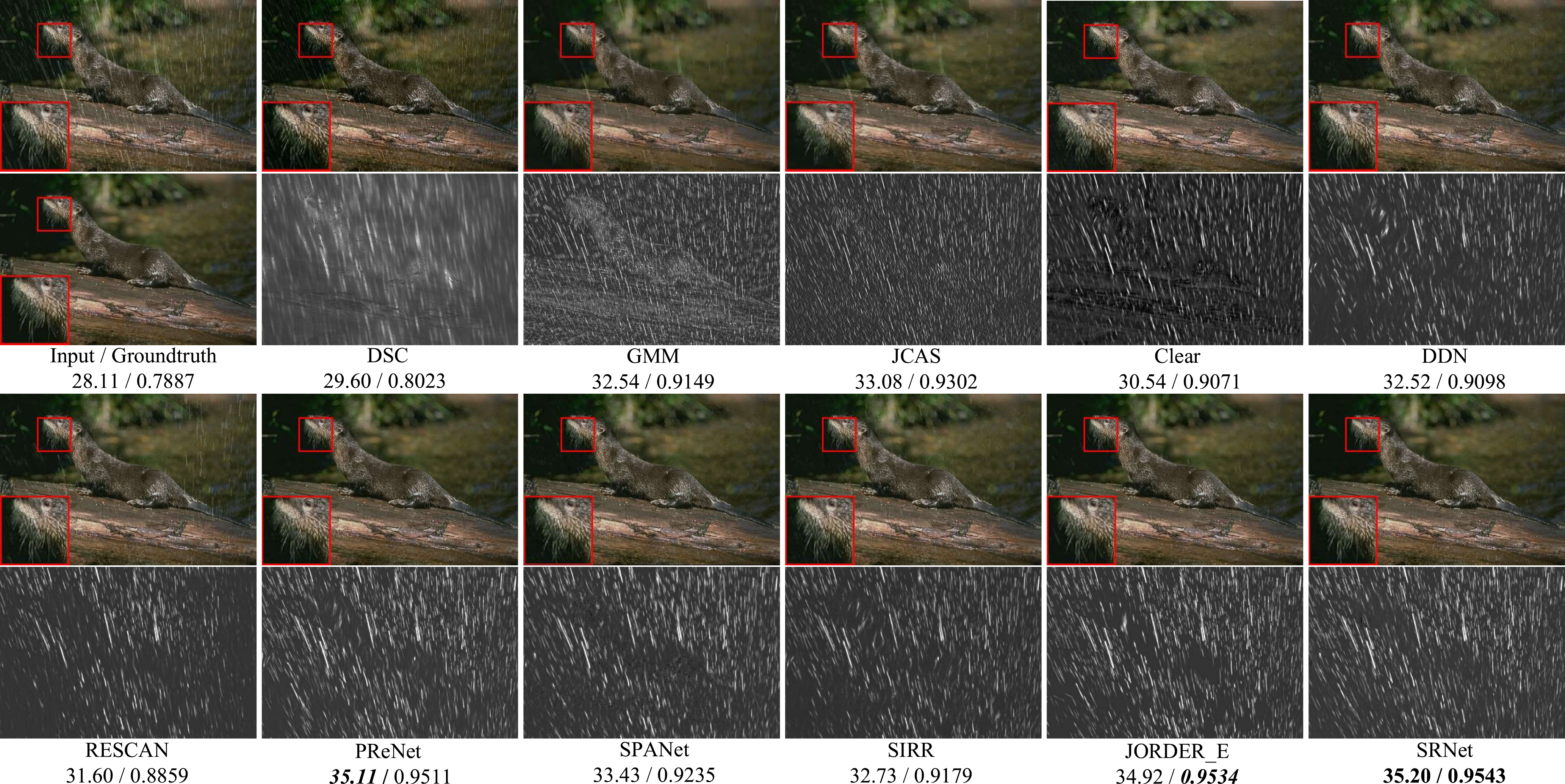}%\vspace{-2mm}
    \caption{Performance comparison of all competing methods on a typical test image in Rain12. The first row: rainy image and derained results obtained by different methods. The second row: groundtruth and rain layers extracted by different methods.}
   % \vspace{-4mm}
    \label{rain12}
\end{figure*}
\section{Experimental Results}
In this section, the effectiveness of the SRNet is verified via comprehensive experiments implemented on synthetic and real datasets by comparing with current state-of-the-art methods, including model based ones:
DSC~\cite{Yu2015Removing}\footnote{\url{https://sites.google.com/view/taixiangjiang/\%E9\%A6\%96\%E9\%A1\%B5/state-of-the-art-methods}},
GMM~\cite{Li2016Rain}\footnote{\url{http://yu-li.github.io/}},
JCAS~\cite{Gu2017Joint}\footnote{\url{https://sites.google.com/site/shuhanggu/home}};
and DL-based ones: Clear~\cite{Fu2017Clearing}\footnote{\url{https://xueyangfu.github.io/projects/tip2017.html}},
DDN~\cite{Fu2017Removing}\footnote{\url{https://xueyangfu.github.io/projects/cvpr2017.html}},
%Zhang \emph{et al.}\cite{did}\footnote{https://github.com/hezhangsprinter/DID-MDN} (denoted as DID-MDN),
RESCAN~\cite{li2018recurrent}\footnote{\url{https://github.com/XiaLiPKU/RESCAN}},
PReNet~\cite{ren2019progressive}\footnote{\url{https://github.com/csdwren/PReNet}},
SPANet~\cite{wang2019spatial}\footnote{\url{https://stevewongv.github.io/derain-project.html}},
JORDER\_E~\cite{Yang2019Joint}\footnote{\url{https://github.com/flyywh}}, SIRR~\cite{wei2019semi}\footnote{\url{https://github.com/wwzjer/Semi-supervised-IRR}}.
\subsection{Experiments on Synthetic Data}
\textbf{Synthetic Datasets.}
We adopt four benchmark datasets: Rain100L~\cite{Yang2019Joint}, Rain100H~\cite{Yang2019Joint}, Rain1400~\cite{Fu2017Removing}, and Rain12~\cite{Li2016Rain}. Rain100L has only one type of rain streaks and consists of 200 image pairs for training and 100 ones for evaluation. With five types of rain streak directions, Rain100H is more challenging and contains 1800 image pairs for training and 100 ones for testing. Rain1400 contains 14000 rainy images synthesized from 1000 clean images with 14 kinds of rain streaks, where 12600 rainy images are used as training samples and 1400 ones as testing samples. Like \cite{ren2019progressive}, the trained model for Rain100L is utilized to evaluate Rain12 that only includes 12 image pairs. For SIRR, we adopt the real 147 rainy images~\cite{wei2019semi} as unsupervised training samples.

\textbf{Performance Metrics.}
Since the groundtruths of synthetic datasets are available, we provide quantitative comparisons based on two commonly used metrics: peak-signal-to-noise ratio (PSNR)~\cite{Huynh2008Scope} and SSIM.
As the human visual system is sensitive to the Y channel of a color image in YCbCr space, we compute PSNR/SSIM based on the luminance channel, and the Matlab computation codes are released by ~\cite{Yang2019Joint}.

\textbf{Evaluation on Rain Removal.}
From the derained results shown in Fig.~\ref{100l}, it is easy to see that traditional model-based DSC, GMM, and JCAS methods leave distinct rain streaks, and DL-based Clear, DDN, and SIRR also leave evident rain marks. Besides, apart from RESCAN, PReNet, and the proposed SRNet, other competing methods blur image textures obviously. From the comparison of extracted rain layers, it is observed that the result by the SRNet covers less texture. Fig.~\ref{rain12} further compares the deraining results on another test image with diverse rain patterns. As displayed, among all comparison methods, SRNet more evidently removes rain streaks and restores background image.

\begin{figure*}[t]
    \centering
        \includegraphics[width=1\linewidth]{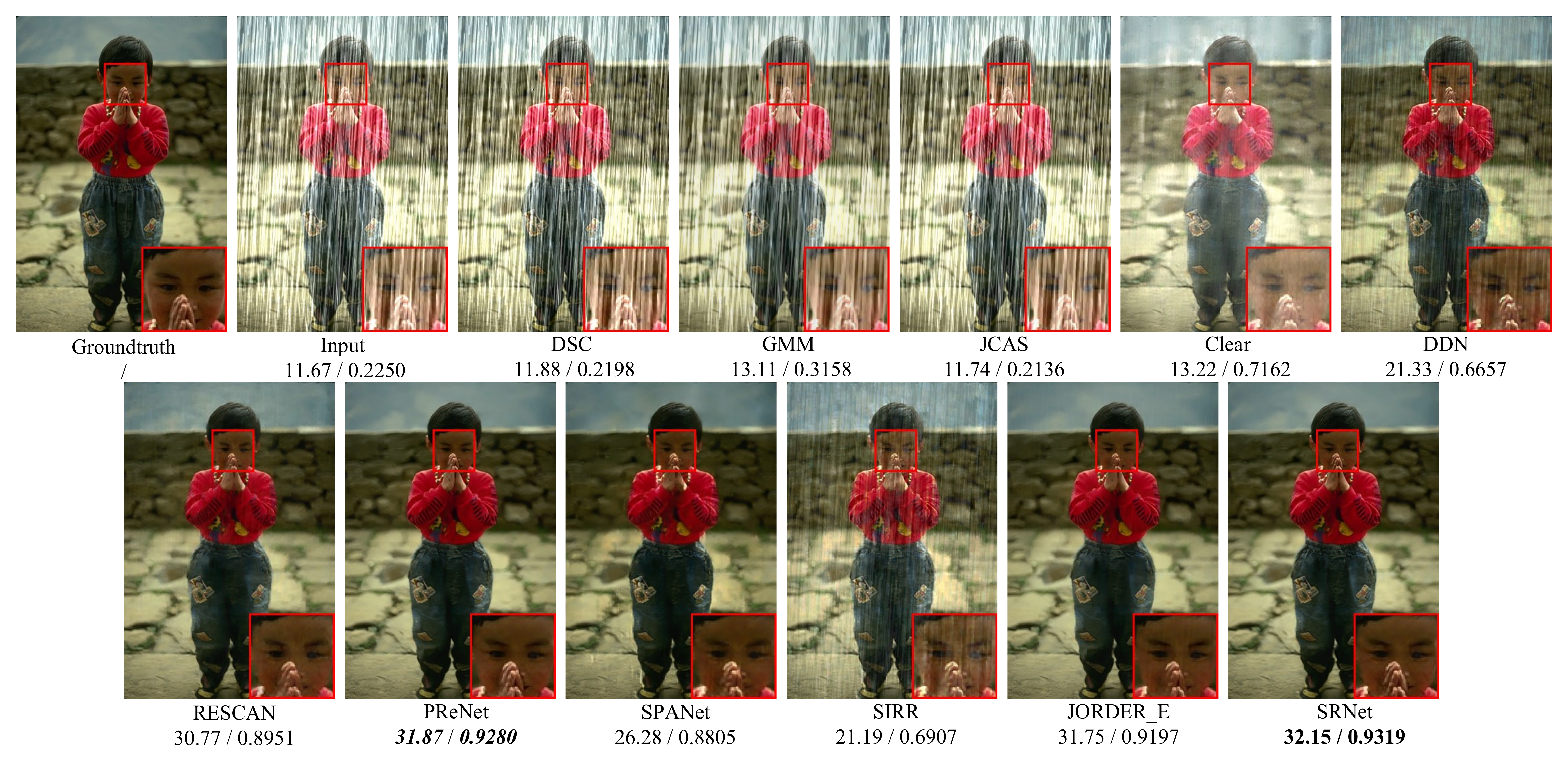}\vspace{-2mm}
        %  \vspace{-2mm}
    \caption{Rain removal effect comparison of all competing methods on a typical test image from Rain100H. .}
    \label{100h}
     \vspace{-2mm}
\end{figure*}
\begin{figure*}[t]
    \centering
        \includegraphics[width=1\linewidth]{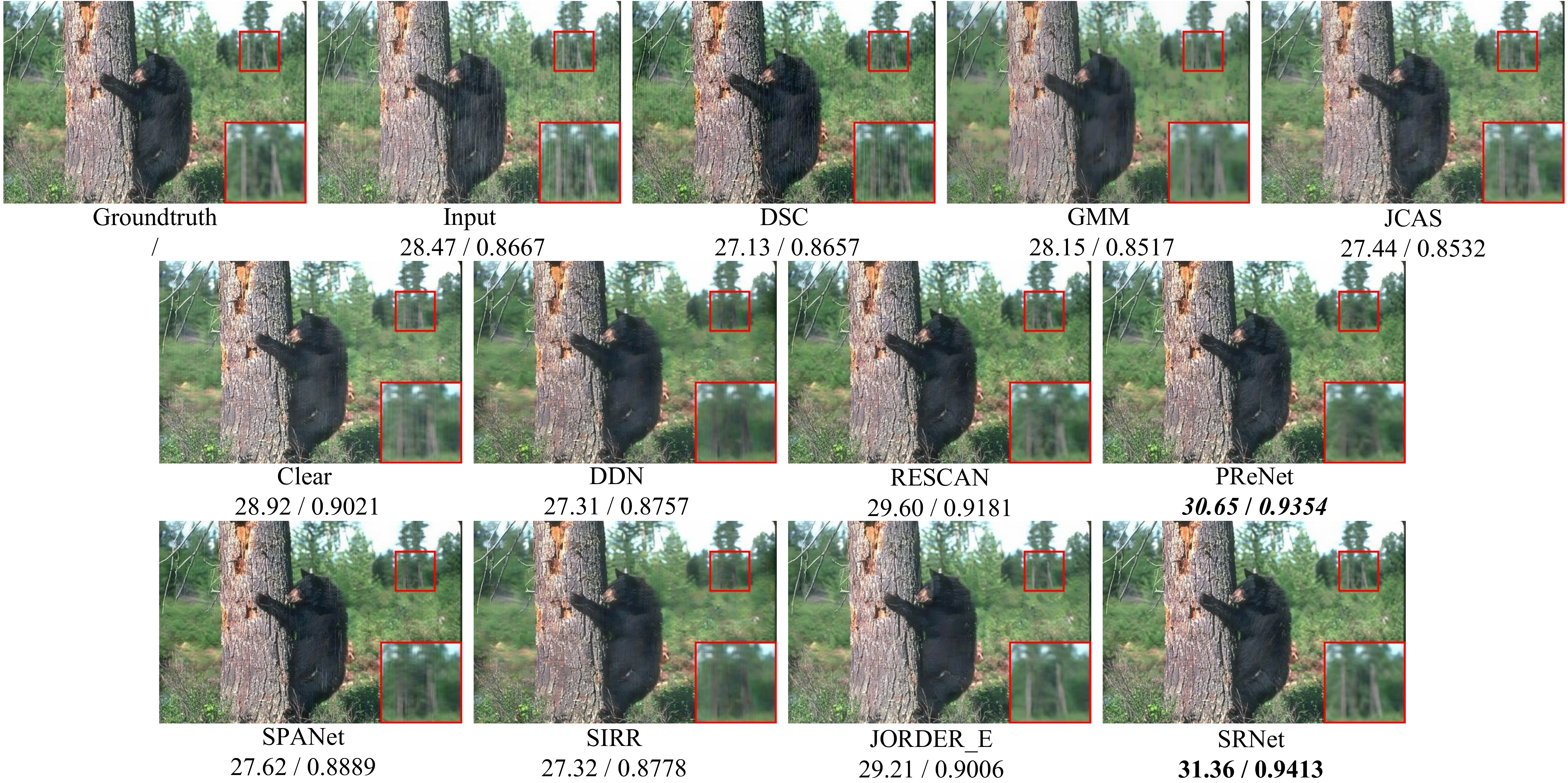}%\vspace{-2mm}
    \caption{Performance comparison of all competing methods on a typical test image in Rain1400.}
    \label{rain1400}
      \vspace{-2mm}
\end{figure*}
\begin{figure*}[t]
    \centering
        \includegraphics[width=1\linewidth]{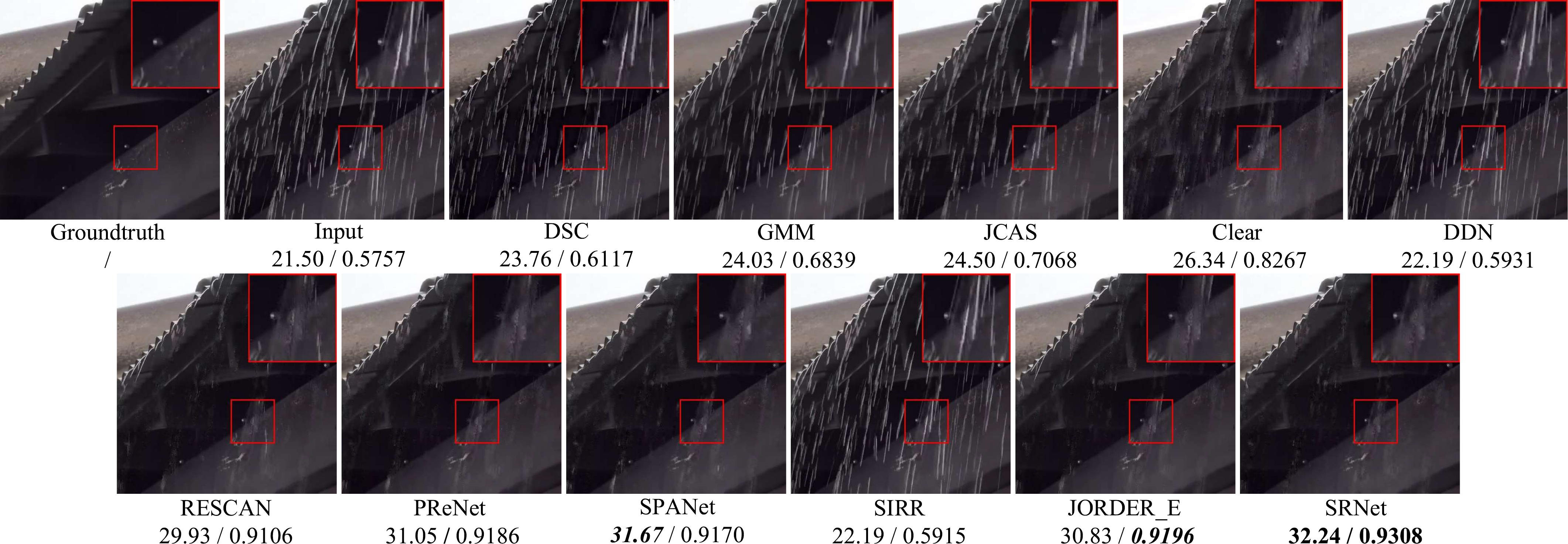}%\vspace{-1mm}
    \caption{Generalization performance comparison of all competing methods on a typical real test image from SPA-Data.}
    \label{wang}
   % \vspace{-1mm}
\end{figure*}
\begin{figure*}[t]
    \centering
        \includegraphics[width=1\linewidth]{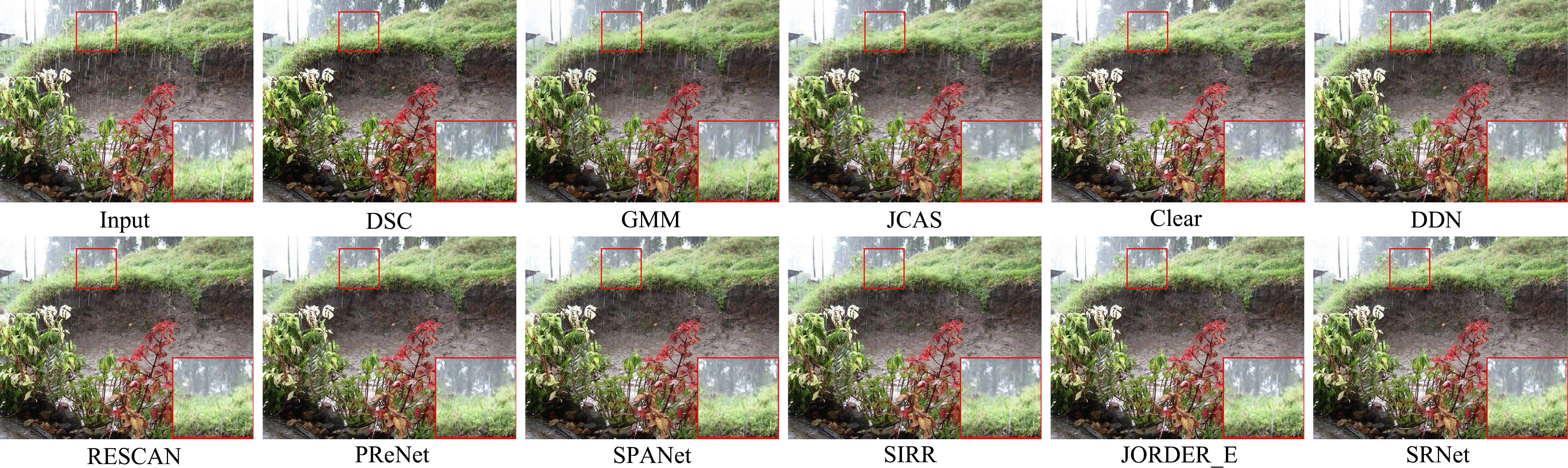}\vspace{-1mm}
    \caption{Generalization performance comparison of all competing methods on a typical real test image with short/light rain streaks from Internet-Data.}
    \label{weil}
    %\vspace{-2mm}
\end{figure*}
\begin{figure*}[t]
    \centering
        \includegraphics[width=1\linewidth]{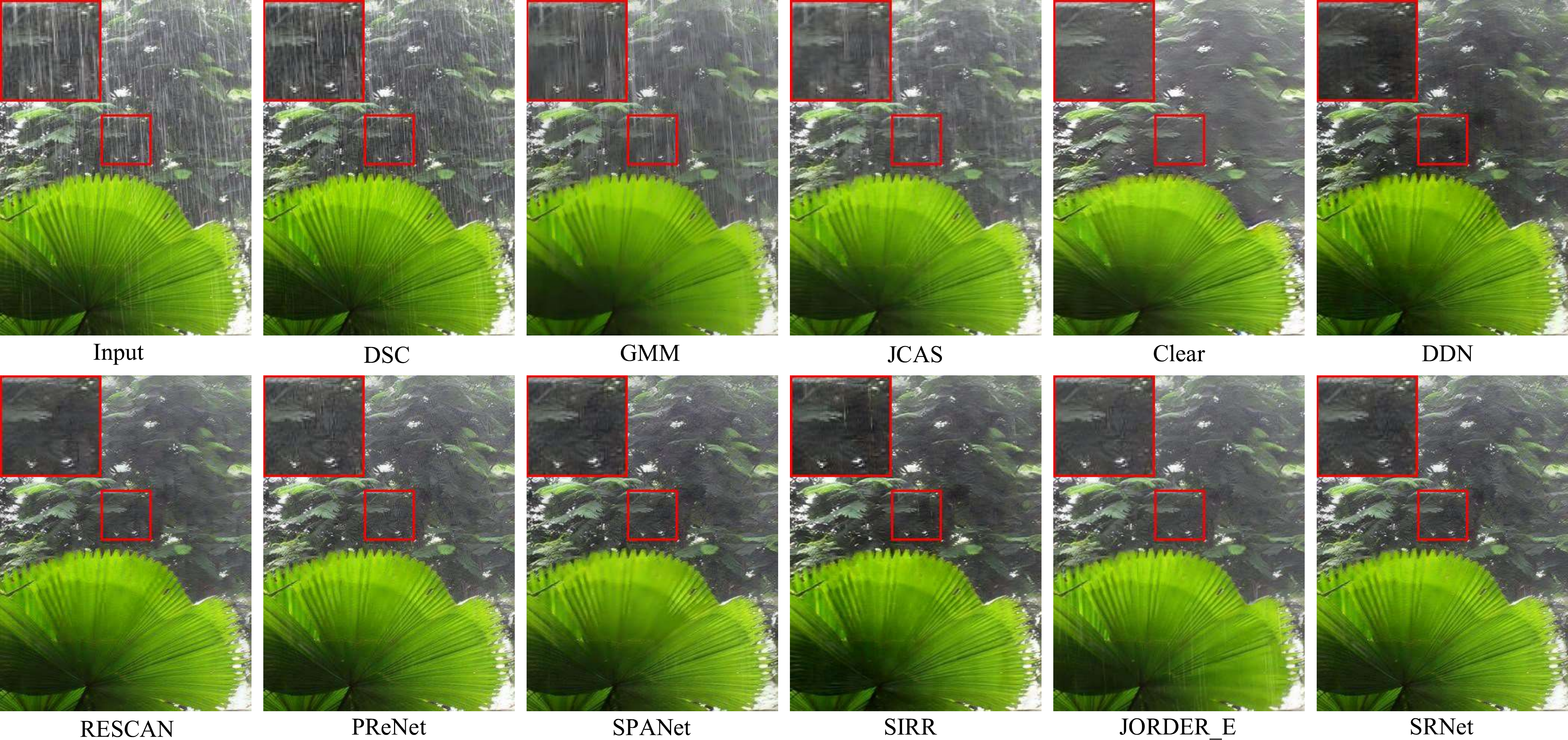}%\vspace{-2mm}
    \caption{Generalization performance comparison of all competing methods on a typical real test image with long/heavy rain streaks from Internet-Data.}
    \label{weih}
        %\vspace{-4mm}
\end{figure*}
\textbf{Evaluation on Detail Preservation.}
We select two more difficult samples from Rain100H and Rain1400, respectively, to evaluate the deraining ability of different methods. As depicted in Fig.~\ref{100h} and Fig.~\ref{rain1400}, complicated rain patterns adversely degrade the rain removal performance of most methods. However, the proposed SRNet still has advantages over rain removal and detail preservation, and achieves best PSNR and SSIM scores.

Tables~\ref{tabsyn} reports the quantitative comparison results of all competing methods on synthetic datasets with diverse, complicated, and different rain types. It is clear that due to the strong fitting ability of CNN, the deraining performance of DL based methods generally outperforms those by conventional prior-model-based methods. Comparatively, the proposed SRNet achieves the best average PSNR and SSIM values on Rain1400 with 14 kinds of rain types. This reflects the better robustness and generality of SRNet.
\begin{table}[t]
%\vspace{-1mm}
\centering
\caption{PSNR and SSIM comparisons on synthetic datasets. Bold and bold italic indicate the best $1^{\text{st}}$ and $2^{\text{nd}}$ performance, respectively.}
% $^{\text{*}}$ means we adopt the pre-trained model released by the authors \cite{wei2019semi}.
%\begin{tabular}{@{}c|c@{}c|c|c|c|c|c|c@{}}
\footnotesize
\setlength{\tabcolsep}{1.1pt}
\begin{tabular}{c|cc|cc|cc|cc}
\Xhline{1.2pt}
  Datasets & \multicolumn{2}{|c|}{Rain100L} & \multicolumn{2}{|c@{}}{Rain100H} & \multicolumn{2}{|c|}{Rain1400} & \multicolumn{2}{|c@{}}{Rain12}\\%& Fig.\ref{e1} & Fig.\ref{e2}& Fig.\ref{e1} &
\Xhline{0.5pt}
  Metrics & PSNR & SSIM & PSNR & SSIM  & PSNR & SSIM & PSNR & SSIM\\
\Xhline{1.2pt}
  Input & 26.90 & 0.8384 & 13.56 & 0.3709 & 25.24 & 0.8097 & 30.14 & 0.8555\\
\Xhline{0.5pt}
  DSC\cite{Yu2015Removing} & 27.34 & 0.8494 & 13.77 & 0.3199  & 27.88 &0.8394 & 30.07 &0.8664\\
\Xhline{0.5pt}
  GMM\cite{Li2016Rain} &29.05 &0.8717 & 15.23 &0.4498  &27.78 & 0.8585 & 32.14 & 0.9145 \\
\Xhline{0.5pt}
  JCAS\cite{Gu2017Joint}  & 28.54 & 0.8524 & 14.62 & 0.4510 &26.20 & 0.8471 & 33.10 &0.9305 \\
\Xhline{0.5pt}
  Clear\cite{Fu2017Clearing} &30.24 & 0.9344 & 15.33 & 0.7421 & 26.21& 0.8951 & 31.24 & 0.9353\\
\Xhline{0.5pt}
  DDN\cite{Fu2017Removing}& 32.38 & 0.9258 & 22.85 & 0.7250 & 28.45 & 0.8888 & 34.04 & 0.9330 \\
%\Xhline{0.5pt}
%  JORDER\cite{Yang2019Joint}$^{\text{o}}$ (CVPR'17) & 36.61 & 0.9740 & 26.54 & 0.8350 \\
\Xhline{0.5pt}
  RESCAN\cite{li2018recurrent}   & \textit{\textbf{38.52}}&\textit{\textbf{ 0.9812}} &29.62 & 0.8720 &32.03& 0.9314 &36.43&0.9519\\
\Xhline{0.5pt}
  PReNet\cite{ren2019progressive}& 37.45& 0.9790 &\textit{\textbf{30.11}}& \textit{\textbf{0.9053}} & \textit{\textbf{32.55}} & \emph{\textbf{0.9459}}& 36.66& 0.9610\\
\Xhline{0.5pt}
  SPANet\cite{wang2019spatial} & 35.33 & 0.9694 &25.11 & 0.8332 & 29.85& 0.9148 & 35.85& 0.9572 \\
\hline
  SIRR\cite{wei2019semi} & 32.37 & 0.9258 & 22.47 & 0.7164 & 28.44 & 0.8893 & 34.02& 0.9347\\
\Xhline{0.5pt}
  JORDER\_E\cite{Yang2019Joint} &\textbf{38.56}&\textbf{0.9827}& \textbf{30.50} &0.8967 &32.00 & 0.9347 &\emph{\textbf{36.69 }}&\emph{\textbf{0.9618}}\\
\Xhline{0.5pt}
SRNet & 37.37 & 0.9780 & 30.05 & \textbf{0.9060} & \textbf{32.88} & \textbf{0.9487} & {\textbf{36.71}} &{\textbf{0.9626}}\\
\Xhline{1.2pt}
\end{tabular}
\label{tabsyn}
\vspace{-3mm}
\end{table}
\begin{table}[t]
%\vspace{2mm}
\centering
\caption{PSNR and SSIM generalization comparisons on SPA-Data~\cite{wang2019spatial}.}
%\begin{tabular}{@{}c|c@{}c|c|c|c|c|c|c@{}}
\footnotesize
\setlength{\tabcolsep}{3.3pt}
\begin{tabular}{p{1cm}<{\centering}|p{1cm}<{\centering}p{1cm}<{\centering}p{0.9cm}<{\centering}p{0.9cm}<{\centering}p{1.3cm}<{\centering}p{1cm}<{\centering}}
\Xhline{1.2pt}
 Methods &Input  &DSC  &GMM &JCAS &Clear &DDN\\%& Fig.\ref{e1} & Fig.\ref{e2}& Fig.\ref{e1} &
\Xhline{1.2pt}
 PSNR &34.15     &34.95   &34.30   &34.95  &32.66  &34.70  \\
\Xhline{0.5pt}
 SSIM &0.9269    &0.9416  &0.9428 &\textbf{0.9453}   &0.9420  &0.9343 \\
\Xhline{1.2pt}
 Methods & RESCAN &PReNet  &SPANet &SIRR &JORDER\_E    &SRNet\\
\Xhline{1.2pt}
 PSNR &34.70    &35.08     &\textit{\textbf{35.13}}    & 34.85  &34.34   &{\textbf{35.31}}  \\
\Xhline{0.5pt}
 SSIM &0.9376   &0.9424  &0.9443  &0.9357   &0.9382  &\textit{\textbf{{0.9448}}} \\
\Xhline{1.2pt}
\end{tabular}
\label{tabwang}
\vspace{-3mm}
\end{table}
\begin{figure}[t]
    \centering
        \includegraphics[width=1\linewidth]{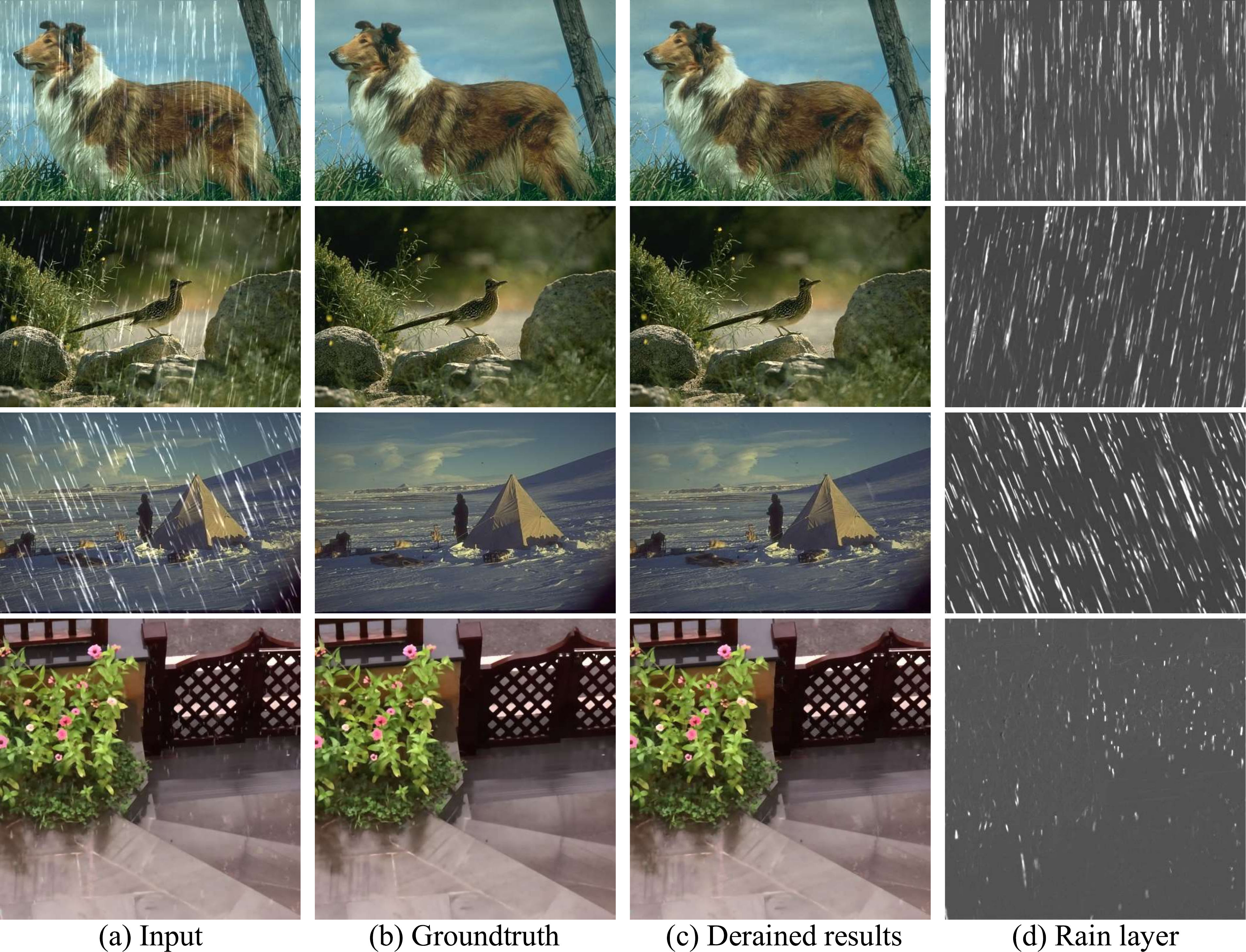}%\vspace{-2mm}
    \caption{(a) Two typical rainy images with different rain types. (b) The corresponding groundtruths. (c) and (d) Derained results and extracted rain layers by SRNet, respectively.}
    \label{rain}
      \vspace{-4mm}
\end{figure}
\begin{figure*}[t]
    \centering
        \includegraphics[width=1\linewidth]{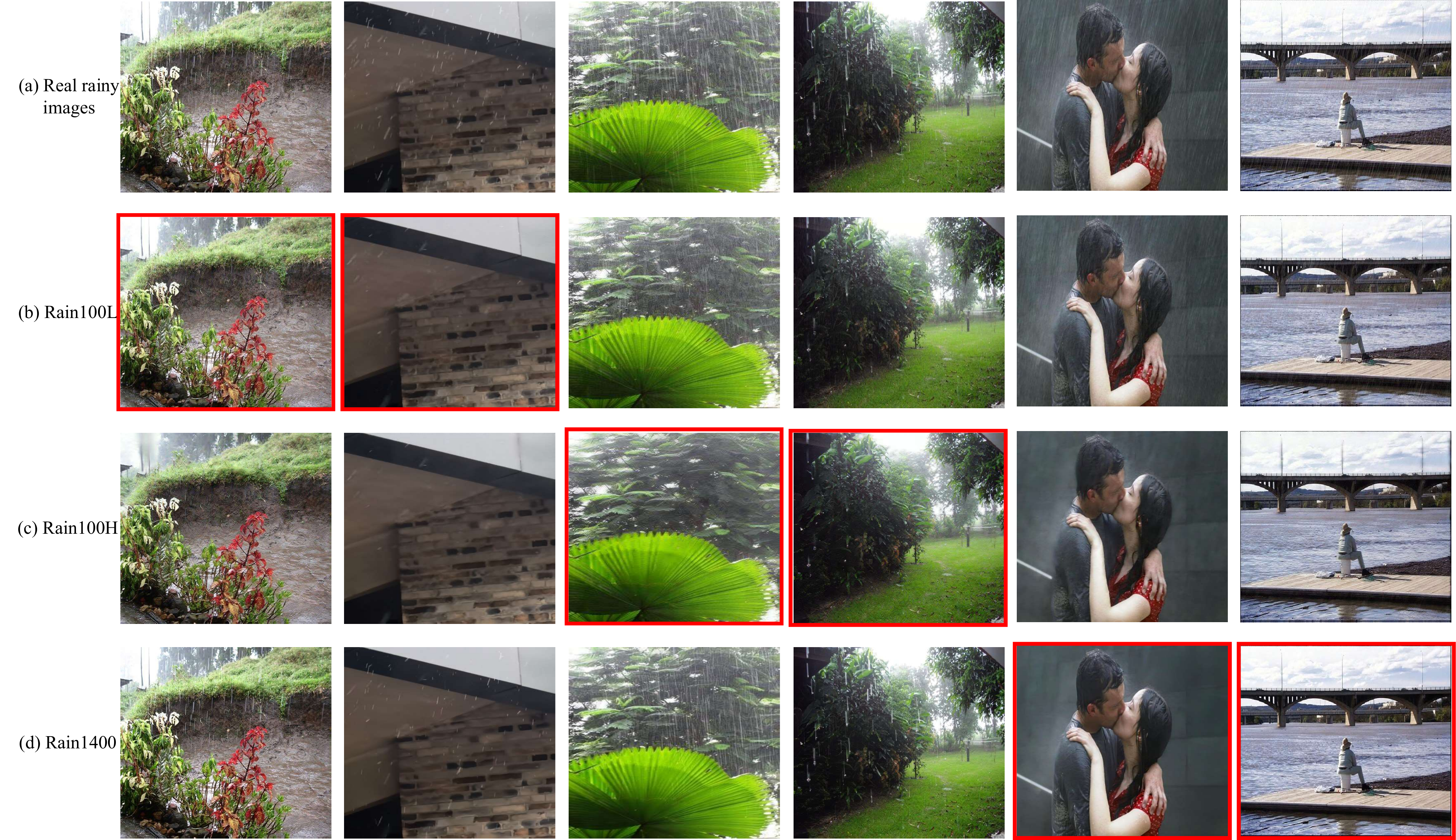}%\vspace{-1mm}
    \caption{Generalization performance comparison of different DL models trained on the corresponding synthetic datasets. (a) Real rainy images. (b)-(d) Derained results predicted by SRNet trained on Rain100L, Rain100H, Rain1400, respectively. Bounded boxes indicate the best ones visually.}
    \label{model_effect}
        %\vspace{-4mm}
\end{figure*}
\subsection{Experiments on Real Data}
\textbf{Real Datasets and Performance Metrics.}
For real application, evaluating the generalization ability of a single image derainer is the key point. In this section, we evaluate the deraining performance of all competing methods on real rainy images, based on two real datasets from~\cite{wang2019spatial} and ~\cite{wei2019semi}, respectively. The former is denoted as SPA-Data and contains 1000 image pairs for evaluation, and the other is called Internet-Data and consists of 147 rainy images without groundtruths. For the experiments on SPA-Data, we compute PSNR and SSIM based on the Y channel. While for Internet-Data, we can only provide visual comparison. Specifically, for each one among the DL-based comparison methods, we can obtain three pre-trained models based on the three synthetic datasets: Rain100L, Rain100H, and Rain1400, respectively. For each method, we select the model with the best PSNR/SSIM on SPA-Data to predict the generalization results for real rainy images from SPA-Data. While for Internet-Data, like previous works, the rain-removed result is the one with the best visual quality among the predicted results by the three models.

\textbf{Evaluation on SPA-Data.}
Fig.~\ref{wang} shows the derained results of all comparison methods on a real rainy image from SPA-Data. As seen, when dealing with the images with complex rain shapes, traditional model based DSC, GMM, and JCAS do not work well, which can be explained by that the adopted specific prior assumptions cannot always accurately represent the complicated rain distribution. DL based competing methods also leave obvious rain streaks and even blur image details, such as PReNet. Nevertheless, as compared with other methods, the proposed SRNet achieve better generalization performance.

To objectively evaluate the generalization ability of these methods, we provide the quantitative rain removal performance comparison on SPA-Data listed in Table~\ref{tabwang}. As reported, the most competitive baseline methods on synthetic datasets, such as RESCAN and JORDER\_E, cannot perform as well as the traditional model-based JCAS on the real dataset, possibly due to the overfitting-the-training-samples issue. It is worth mentioning that in this case, JCAS obtains a good SSIM as the method utilizes the complementary of ASR and SSR to finely describe texture structures. Although SPANet has relatively inferior deraining effect on synthetic datasets, its generalization performance is competitive. Observing the comparison results presented in Table~\ref{tabsyn} and Table~\ref{tabwang},  it is easy to see that among all comparison methods, the proposed SRNet is a competing single image derainer as it achieves satisfactory rain removal performance on both synthetic and real datasets. This confirms the effectiveness of structural residual design of our network architecture and the fine generalization ability of SRNet.
%\begin{table}[t]
%\centering
%\caption{PSNR and SSIM comparisons on SPA-Data\cite{wang2019spatial}. Bold and bold italic are used to indicate top $1^{\text{st}}$ and $2^{\text{nd}}$ rank, respectively.}
%%\begin{tabular}{@{}c|c@{}c|c|c|c|c|c|c@{}}
%\begin{tabular}{@{}C{1.8cm}@{}|@{}C{1.2cm}@{}@{}C{1.2cm}@{}|@{}C{2.1cm}@{}|@{}C{1.2cm}@{}@{}C{1.2cm}@{}@{}}
%  \hline
%  % after \\: \hline or \cline{col1-col2} \cline{col3-col4} ...
%\Xhline{1.2pt}
% Methods & PSNR  & SSIM & Methods & PSNR  & SSIM\\%& Fig.\ref{e1} & Fig.\ref{e2}& Fig.\ref{e1} &
%\Xhline{1.2pt}
% Input & 34.15  & 0.9269  & RESCAN\cite{li2018recurrent}  & 34.70  &0.9376 \\
%\Xhline{0.5pt}
%DSC\cite{Yu2015Removing} & 34.95  & 0.9416 &PReNet\cite{ren2019progressive} & 35.08  &0.9424  \\
%\Xhline{0.5pt}
%GMM\cite{Li2016Rain}  &34.30  & 0.9428 & SPANet\cite{wang2019spatial}  &\emph{\textbf{35.13}} & 0.9443 \\%& Fig.\ref{e1} & Fig.\ref{e2}& Fig.\ref{e1} &
%\Xhline{0.5pt}
%JCAS\cite{Gu2017Joint}  & 34.95 &\textbf{0.9453} & SIRR \cite{wei2019semi} & 34.85   & 0.9357  \\
%\Xhline{0.5pt}
%Clear\cite{Fu2017Clearing} & 32.66  & 0.9420 &  JORDER\_E\cite{Yang2019Joint} & 34.34 &0.9382\\
%\Xhline{0.5pt}
%DDN\cite{Fu2017Removing}  & 34.70   &0.9343 & MSEDNet &{\textbf{35.31}} &\emph{\textbf{{0.9448}}}\\
%\Xhline{1.2pt}
%\end{tabular}
%\label{t3}
%\end{table}

\textbf{Evaluation on Internet-Data.}
Furthermore, we select another two hard samples from Internet-Data with complicated rain distribution. Fig.~\ref{weil} and Fig.~\ref{weih} demonstrate the corresponding generalization comparison results. From these figures, similar to the experiment analysis on SPA-Data, traditional model-based methods always leave severe rains, and other DL based methods still have obvious rain marks and blur image details. Comparatively, the proposed SRNet has evident superior performance in rain removal and detail preservation.

\subsection{Rain Layer Extraction Evaluation on SRNet}
Here we select several rainy images with different rain shapes from aforementioned datasets to further evaluate the effectiveness of the proposed SRNet on extracting rain layer. As shown in Fig.~\ref{rain}, when dealing with complicated rainy images with diverse rain streaks (light/heavy density, short/long shape, different directions), the proposed method can finely extract rain layer and preserve background details well.
\subsection{Ablation Studies}
\textbf{The Effect of Training Datasets.}
Based on some typical real rainy images, Fig.~\ref{model_effect} shows the derained results predicted by SRNet trained on the three synthesized datasets, including Rain100L, Rain100H, and Rain1400. For each input rainy image, we use a box to indicate the best recovery result visually. It can be observed that the model trained on Rain100L performs well in the case of light/thin rain streaks; the trained model on Rain100H is applicable to heavy/accumulated rain types, and for Rain1400, it is suitable for long and thin ones.

\textbf{Network Modules Analysis.}
Here we quantitatively analyze the importance of different modules of the proposed SRNet, as listed in Table~\ref{t6}. We train all the variants on Rain100L and compare their generalization ability on SPA-Data. $B_{a}$, $B_{b}$, $B_{c}$, and $B_{d}$ denote four variants of only using the small scale branch (DF=1) in Fig.~\ref{mseda}(a). $B_{a}$ and $B_{b}$ are basic Resnet-like networks, only consisting of Resblocks. $B_{c}$ and $B_{d}$ include symmetrical downsampling and unsampling layers following the corresponding Resblocks. $B_{f}$ is the default SRNet. $B_{e}$ is a variant that SRNet adopts weight sharing among three parallel branches, which shows that the SRNet can notably reduce network parameters by weight sharing without unsubstantial degradation in generalization performance. It is worth mentioning that even $B_{d}$ only adopts small scale, it outperforms most other competing methods on real deraining performance as shown in Table~\ref{tabwang}, which corroborates the rationality of the proposed encoder-decoder network. The versus between every two variants and the corresponding functioning module is concluded as Table~\ref{t7}.
\begin{table}[t]%\footnotesize
\centering
\caption{Network module analysis of the proposed SRNet.}
%\resizebox{\textwidth}{8mm}{
%\tabcolsep 0.001mm
\setlength{\tabcolsep}{2.4pt}
%\begin{tabular}{@{}c|c@{}c|c|c|c|c|c|c@{}}
\begin{tabular}{c|c|cccccc}
  \hline
\Xhline{1.2pt}
\multicolumn{2}{c|} {Methods} & $B_{a}$  & $B_{b}$& $B_{c}$ & $B_{d}$ & $B_{e}$ & $B_{f}$\\%& Fig.\ref{e1} & Fig.\ref{e2}& Fig.\ref{e1} &
\Xhline{1.2pt}
%\multicolumn{2}{c|} {Single scale} & $\checkmark$ & $\checkmark$ & $\checkmark$ &$\checkmark$  & & \\
%\Xhline{0.5pt}
\multicolumn{2}{c|} {Multi-Scale}& \quad& & & & $\checkmark$ & $\checkmark$ \\
\Xhline{0.5pt}
\multicolumn{2}{c|} {Resblock w/o pooling} &$\checkmark$  & $\checkmark$ & & & & \\
\Xhline{0.5pt}
\multicolumn{2}{c|} {Encoder-Decoder Net} &\quad & &$\checkmark$  & $\checkmark$ & $\checkmark$ &$\checkmark$  \\
\Xhline{0.5pt}
\multicolumn{2}{c|} {Bilinear unsampling}  & \quad& & $\checkmark$ & & & \\
\Xhline{0.5pt}
\multicolumn{2}{c|} {MaxUnpooling} & \quad& & & $\checkmark$ & $\checkmark$ & $\checkmark$ \\
\Xhline{0.5pt}
\multicolumn{2}{c|}  {Global residual learning}&\quad & $\checkmark$ &$\checkmark$  &$\checkmark$  &$\checkmark$  & $\checkmark$ \\
\Xhline{0.5pt}
\multicolumn{2}{c|}  {Weight sharing} &\quad & & & & $\checkmark$ & \\
\Xhline{0.5pt}
\multicolumn{2}{c|}  {w/o Weight sharing} &\quad & & & & &$\checkmark$  \\
\Xhline{0.5pt}
%\multirow{2}*{Rain100L} & PSNR    &36.87 & 36.45& \textit{\textbf{37.10}}& 36.90& 36.72  &{\textbf{ 37.37}} \\
%                        & SSIM    &0.9773 & 0.9754& \textit{\textbf{0.9779 }}&0.9772 &0.9759 &\textbf{0.9780}\\
%
%\Xhline{0.5pt}
\multirow{2}*{SPA-Data} & PSNR    &35.06 &35.15& 35.20&\textbf{\textit{35.23}}& 35.22& \textbf{{35.31}}  \\
                            & SSIM    &0.9414& 0.9437& 0.9438& 0.9441 &\textit{\textbf{0.9444}} &{\textbf{0.9448}}\\
\Xhline{1.2pt}
\end{tabular}
\label{t6}
%\vspace{-2mm}
\end{table}
\begin{table}[t]
%\vspace{-1mm}
\centering
\caption{Versus among the variants of SRNet given in Table~\ref{t6}.}
\setlength{\tabcolsep}{18pt}
\begin{tabular}{c|c|c}
\hline
\Xhline{1.2pt}
Versus & Winner & Effective Module\\
\Xhline{1.2pt}
$B_{a}$ vs $B_{b}$  & $B_{b}$ & Global residual learning  \\
\Xhline{0.5pt}
$B_{b}$ vs $B_{c}$  & $B_{c}$& Encoder-Decoder Net\\
\Xhline{0.5pt}
$B_{c}$ vs $B_{d}$ & $B_{d}$ & MaxUpooling   \\
\Xhline{0.5pt}
$B_{d}$ vs $B_{f}$ & $B_{f}$ & Multi-Scale\\
\Xhline{1.2pt}
\end{tabular}
\label{t7}
 %\vspace{-2mm}
\end{table}

\subsection{Extension to Real Video Deraining}
Finally, we evaluate the deraining ability of the SRNet on real rainy videos without groundtruth, by comparing with current state-of-the-art video deraining methods, including the traditional model-based Garg \emph{et al.}~\cite{gk_vision}, Kim \emph{et al.}~\cite{Jin2015Video}, Jiang \emph{et al.}~\cite{jiang}, Ren \emph{et al.}~\cite{Ren2017Video}, Wei \emph{et al.}~\cite{wei2017should}, Li \emph{et al.}~\cite{li2018video}, and the DL-based Liu \emph{et al.}~\cite{liu2018erase}.

\begin{figure*}[htb]
    \centering
        \includegraphics[width=1\linewidth]{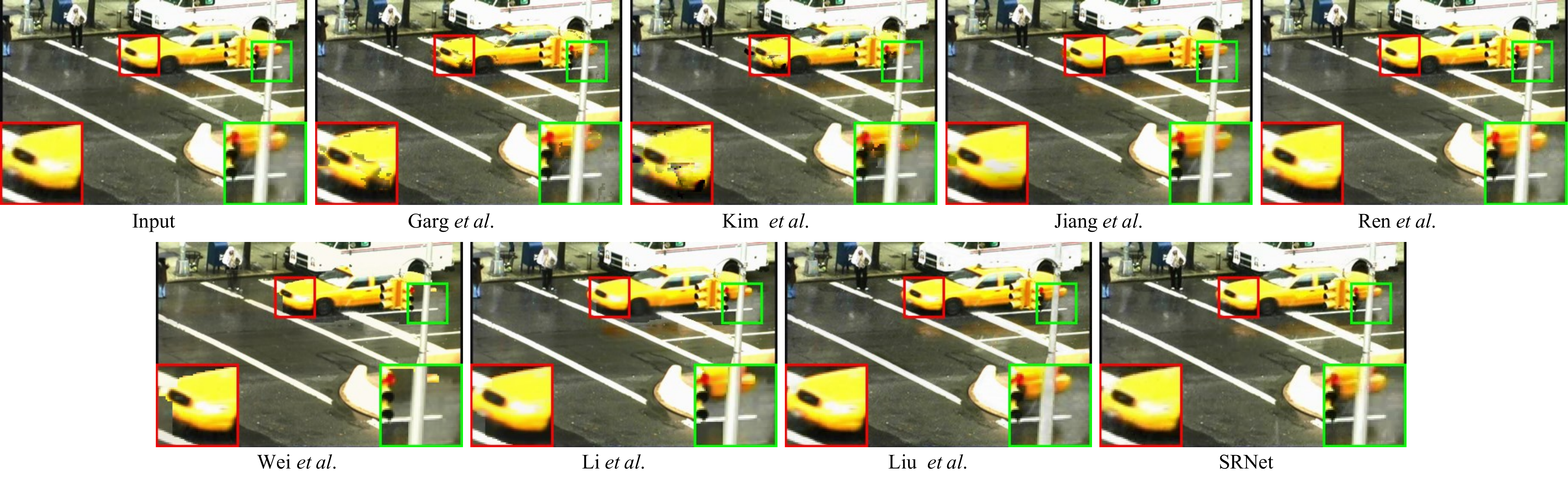}%\vspace{-2mm}
    \caption{Deraining performance of competing video deraining methods and SRNet on a real video frame with complex moving objects.}
    \label{comp}
     % \vspace{-2mm}
\end{figure*}

\begin{figure*}[htb]
    \centering
        \includegraphics[width=1\linewidth]{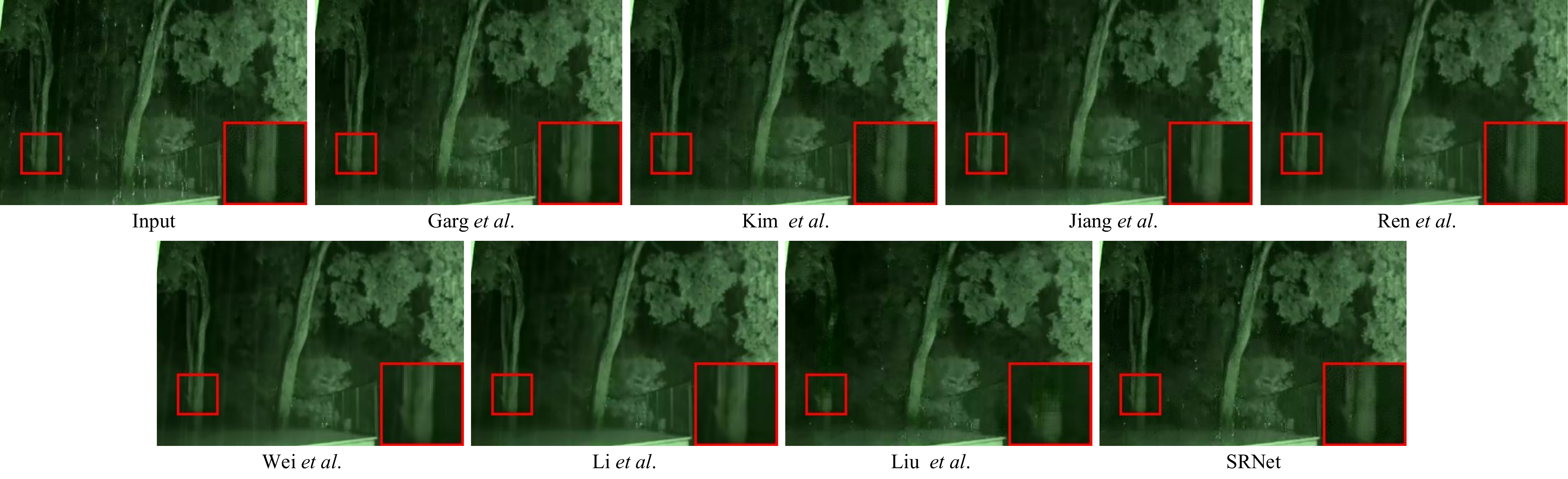}%\vspace{-2mm}
    \caption{Deraining performance of competing video deraining methods and SRNet on a real video frame captured at night.}
    \label{night}
   %  \vspace{-2mm}
\end{figure*}

Fig.~\ref{comp} and Fig. \ref{night} present the visual comparison results based on two real rainy frames: one is captured by surveillance systems on a street with complex moving objects, and the other is obtained at night. From Fig.~\ref{comp}, it is easily seen that the model-based video deraining methods, including Garg \emph{et al.}'s, Kim \emph{et al.}'s, Jiang \emph{et al.}'s, Wei \emph{et al.}'s, and Li \emph{et al.}'s, cause different degrees of artifacts at the location of the moving car, and the DL based Liu \emph{et al.}'s method leaves obvious rain streaks.
Even not utilizing temporal information among multi-frames, the proposed single image deraining method, SRNet, still performs well on removing rain streaks and preserving background textures. From Fig. \ref{night}, SRNet has the ability to detect obvious rain streaks. Although some model-based video deraining methods can remove more rain streaks than SRNet, they need many frames and thus have unfavorable real-time performance. It is noteworthy that as compared with the DL-based Liu \emph{et al.}'s video deraining method, SRNet can also preserve relatively more image details, like trunks.
%\begin{figure}[htb]
%    \centering
%        \includegraphics[scale=0.42]{video.pdf}\vspace{-1mm}
%    \caption{Rain removal effect of MSEDNet on two \textbf{real rainy videos}. (a) Rainy videos. (b) Derained results.}
%    \label{video}
%      \vspace{-3mm}
%\end{figure}
 %  \vspace{-3mm}
\section{Conclusions and Future Work}
In this paper, we have taken into account the intrinsic prior structures of rain streaks, and designed a specific structural residual network for rain layer removal from a single image. By visualizing the extracted rain layer and sparse rain feature maps for synthetic and real test images, we have validated the working mechanism underlying the proposed structural residual network. Comprehensive experiments have demonstrated the superiority of the proposed method over current state-of-the-art single image derainers in terms of both robustness and generalization capability.

Although we have fairly evaluated the generalization performance of all competing methods based on SPA-Data, the dataset relies on human judgments and it cannot cover all the complicated patterns of rains in real-world. Besides, collecting a large number of really real rainy/clean image pairs is extremely time-consuming and cumbersome. To alleviate these problems, in the future, we will make further efforts for employing such network designing idea into unsupervised or at least semi-supervised scenarios in our future investigations.
%
%Besides, from the perspective of real application, it should be more meaningful to train a unified model by mixing part samples selected from different synthetic datasets. One one hand, it can simplify the generalization comparison process. One the other hand, it would improve the current generalization performance due to the diversity of rain patterns during the training phase.

\ifCLASSOPTIONcaptionsoff
  \newpage
\fi

%\bibliography{egbib_tip}
\bibliography{bib_tip}
%\section*{Acknowledgements}
\end{document}